\documentclass{article}
\usepackage{kmn}
\usepackage{epsfig}
\title[Infrared spectroscopy of high excitation planetary nebulae]{The
coronal line regions of planetary nebulae NGC\,6302 and
NGC\,6537: 3--13$\mu$m grating and echelle spectroscopy}

\author[S.~Casassus et al.]
  {S.~Casassus,$^1$ P.\,F.~Roche,$^1$ M.\,J.~Barlow$^2$\\
	$^1$ Astrophysics, Nuclear and Astrophysics Laboratory, Physics Department, Oxford University, Keble Road, Oxford OX1 3RH\\
	$^2$ Department of Physics and Astronomy, University College
London, Gower Street, London WC1E 6BT\\}
\pagerange{\pageref{firstpage}--\pageref{lastpage}}
\pubyear{1998}
\def\gs{\mathrel{\raise1.16pt\hbox{$>$}\kern-7.0pt
\lower3.06pt\hbox{{$\scriptstyle \sim$}}}}
\def\ls{\mathrel{\raise1.16pt\hbox{$<$}\kern-7.0pt
\lower3.06pt\hbox{{$\scriptstyle \sim$}}}}
\begin{document}
\label{firstpage}
\maketitle
\begin{abstract}
We report on advances in the study of the cores of NGC~6302 and
NGC~6537 using infrared grating and echelle spectroscopy. In NGC~6302,
emission lines from species spanning a large range of ionization
potential, and in particular \hbox{[Si\,{\sc ix}]} 3.934~$\mu$m, are
interpreted using photoionization models (including CLOUDY), which
allow us to reestimate the central star's temperature to be about
250\,000~K.  All of the detected lines are consistent with this value,
except for [Al\,{\sc v}] and [Al\,{\sc vi}]. Aluminium is found to be
depleted to one hundredth of the solar abundance, which provides
further evidence for some dust being mixed with the highly ionized gas
(with photons harder than 154~eV). A similar depletion pattern is
observed in NGC~6537. Echelle spectroscopy of IR coronal ions in
NGC~6302 reveals a stratified structure in ionization potential, which
confirms photoionization to be the dominant ionization mechanism. The
lines are narrow ($<$ 22~km\,s$^{-1}$ {\footnotesize{FWHM}}), with no
evidence of the broad wings found in optical lines from species with
similar ionization potentials, such as \hbox{[Ne\,{\sc
v}]}\,3426~\AA. We note the absence of a hot bubble, or a wind blown
bipolar cavity filled with a hot plasma, at least on $1''$ and
10~km\,s$^{-1}$ scales.  The systemic heliocentric velocities for
NGC~6302 and NGC~6537, measured from the echelle spectra of IR
recombination lines, are found to be --34.8$\pm$1~km\,s$^{-1}$ and
--17.8$\pm$3~km\,s$^{-1}$. 
We also provide accurate new wavelengths for
several of the infrared coronal lines observed with the echelle.

\end{abstract}
\section{Introduction}

The extreme high excitation PNe NGC~6302 and NGC~6537, which are both
bipolar and of Peimbert's \mbox{Type I}, show lines from the highest
ionization species found in PNe (e.g. \hbox{[Si\,{\sc vi}]}
1.96~$\mu$m, Ashley \& Hyland 1988, hereafter AH88) as well as fast
flows of order 1000~km\,s$^{-1}$ (Meaburn \& Walsh 1980). Proposed
ionization mechanisms include photoionization by a very hot central
star, or shock excitation (e.g.  Rowlands et al. 1994, Lame \& Ferland
1991, AH88, Barral et al. 1982). Pottasch et al. (1996) published a
portion of the {\it ISO} ({\em Infrared Space Observatory}) spectrum
of NGC~6302, and found that the properties of the spectrum can be well
accounted for by photoionization by a 380\,000~K greybody central
star, using the photoionization code {\small CLOUDY} (Ferland 1996).

We report here on a study of the cores of NGC~6302 and NGC~6537 using
near- and mid-infrared spectroscopy. Low and medium resolution CGS3
and CGS4 infrared spectra taken at UKIRT have higher sensitivity in
the atmospheric windows than the SWS on {\it ISO}, and our spectrum of
NGC~6302 shows the presence of \hbox{[Si\,{\sc ix}]} 3.934$\mu$m,
which had only been tentatively detected in the {\it ISO}
spectrum. This detection has allowed us to reestimate the central
star's temperature to be about 250\,000~K, the upper limit inferred by
Marconi et al. (1996) from the non-detection of \hbox{[Si\,{\sc ix}]}
by Oliva et al. (1996).

We also present unprecedently high dispersion infrared echelle spectra
of the cores of these nebulae. Knowledge of the physical conditions and
velocity
fields in the obscured cores of bipolar PNe is the key to constructing
models, but these are currently inferred only through extrapolation from
optical studies. We show here that tracing the velocity field with
coronal ions leads to quite a different picture of the structure of
these bipolar planetary nebulae.

The infrared coronal lines are in fact very narrow, with no evidence
of the fast flows detected in \hbox{[Ne\,{\sc v}]}~3426\,\AA~ towards
the core of NGC~6302 by Meaburn \& Walsh (1980). We confirm the
inference by AH88 that the full width at zero intensity of
\hbox{[Si\,{\sc vi}]} is less than $\sim$30~km\,s$^{-1}$. The absence
of broad wings in the infrared lines constitutes another piece of
information with which to understand the physical conditions in the
regions where the \hbox{[Ne\,{\sc v}]}~3426\,\AA~ broad wings
originate.

The CGS4 spectra allow the determination of several coronal line
wavelengths, which we measure to be significantly different from the
$ISO$ results (Feuchtgruber et al. 1997). We present a thorough
account of our wavelength calibration procedures, and a discussion of
their accuracy.

We describe our observations in Section \ref{sec:observations}, the
analysis of the spectra and the determination of accurate wavelengths
for several coronal ions are deferred until the Appendix. Section
\ref{sec:abundances} consists of the model-dependent interpretation of
our observations. We present an abundance analysis for NGC~6302 and
NGC~6537 and estimate the central stars' temperatures. In that Section
we also use narrow band images of NGC~6302 to infer properties of the
progenitor. Section \ref{sec:kine} contains an interpretation of the
kinematical information from the echelle results, while
Section~\ref{sec:concl} summarises our conclusions.

\section{Observations}
\label{sec:observations}
\subsection{Low resolution spectroscopy}

\begin{figure}
 \resizebox{8cm}{!}{\epsfig{file=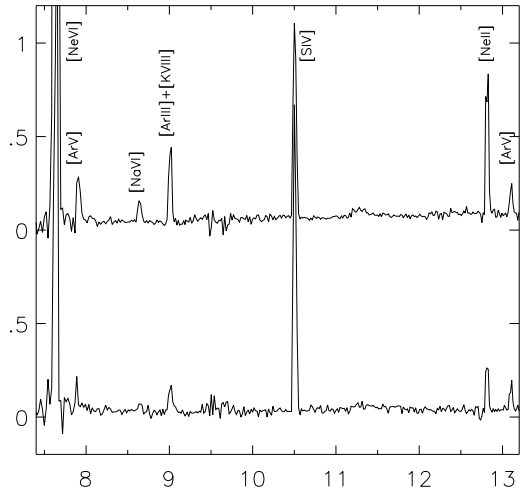}}
 \caption{CGS3 spectra of NGC~6302 and
 NGC~6537. The spectrum with annotations is that of NGC
 6302. Wavelengths are in microns, and flux densities are in
$10^{-12}$~W~m$^{-2}$~$\mu$m$^{-1}$.}  \label{fig:speccgs3}
\end{figure}

The 8-13 $\mu$m spectra (atmospheric N-band window) of NGC~6302,
PN\,G349.5+01.0, and NGC~6537, PN\,G010.1+00.7 were obtained with CGS3
at UKIRT, on May 15 1994. The observations were chopped for background
cancellation. The 32 Si:As detectors and the high resolution grating
provide a resolution of 0.05$\mu$m (R=200 at 10$\mu$m), sampled 3
times per resolution element. The reduced spectra, obtained through a
4$''$ circular aperture, are shown in Figure \ref{fig:speccgs3}.

\begin{figure*}
 \resizebox{\textwidth}{!}{\epsfig{file=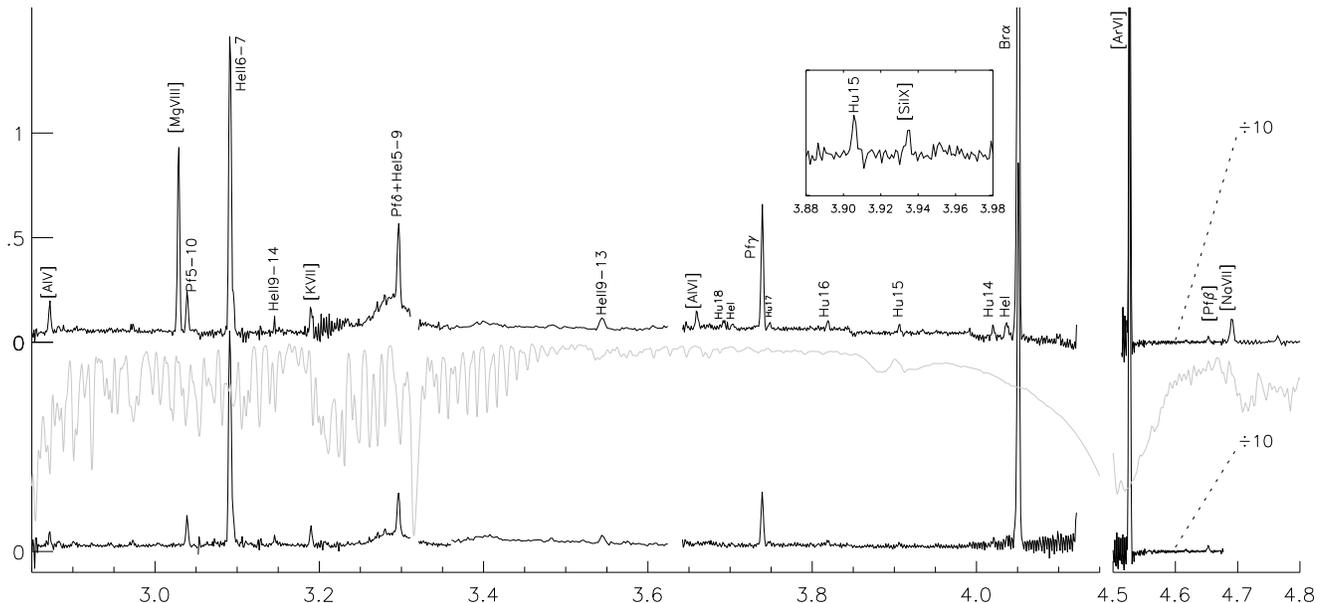}}
 \caption{CGS4 grating spectra of NGC 6302 (upper) and NGC~6537
 (lower). Wavelengths are in microns, and flux densities in
 $10^{-12}$~W~m$^{-2}$~$\mu$m$^{-1}$.  Superposed on NGC~6537's
 spectrum, in light grey, is a low resolution model
 atmospheric transmission function appropriate for the summit of Mauna
 Kea.}  \label{fig:speccgs4}
\end{figure*}

NGC~6302 and NGC~6537 were also observed at UKIRT through the
atmospheric windows centred on the L and M filter band-passes, on
April 24, 25 and 26 1994, using CGS4 equipped with the short
camera. The observations were carried out in chopping mode to allow
background cancellation, at a resolving power of R=1500, a spatial
resolution of 3$''$ along the slit and a slit width of 3$''$, with the
spectra
oversampled three times by shifting the detector's position 6 times
over two pixels. The CGS4 detector at the time was a 62x58 InSb
array. Due to problems with the array electronics producing random bad
pixels in the first 20 columns of the array, the spectra were
extracted using the minimum number of rows necessary to cover most of
the emission from the nebula. The reduced spectra are shown in Figure
\ref{fig:speccgs4}.  The dominant features are fine structure emission
lines from high stages of ionization, which probe the cores of the
nebulae.

\subsection{Echelle spectroscopy.}

The nebulae were observed with the CGS4 short camera and echelle at
UKIRT on May 11 and 12 1997. Two out of three nights were clear, so
the observations are not as complete as hoped. In the case of
NGC~6302, an infrared disk lies at the waist of the bipolar structure,
where a dark lane is seen in optical images (Lester \& Dinerstein
1984).  We oriented the CGS4 slit along two perpendicular position
angles, the $\parallel$ position lies at $-20^{\circ}$, along the
dark lane at the waist of the nebula, and the $\perp$ position lies at
--110$^{\circ}$, along the bipolar axis. We only obtained echelle
spectra with the $\parallel$ position angle for NGC~6537, at
--45$^{\circ}$.  The list of observed lines and their orientation in
NGC~6302 is: Br$\gamma$, $\perp$ \& $\parallel$; \hbox{[Mg\,{\sc
viii}]}, $\perp$ \& $\parallel$; \hbox{[Si\,{\sc vi}]}, $\parallel$;
\hbox{[Ar\,{\sc vi}]}, $\perp$. In NGC~6537, we observed Br$\gamma$,
\hbox{[Si\,{\sc vi}]}, and \hbox{[Al\,{\sc v}]}, all at the
$\parallel$ slit orientation.

At wavelengths near 4$\mu$m, the order sorting CVF filter produced
interference fringes, hampering the data acquisition. In the case of
\hbox{[Ar\,{\sc vi}]}, the line was close to a fringe maximum, and the
spatial distribution could still be determined. This ionic line was
one of the best candidates to trace the velocity field since it is one
of the very brightest, and also because its ionization potential is
only a little less than that of \hbox{Ne\,{\sc v}}, in which Meaburn
\& Walsh (1980) detected 1000~km\,s$^{-1}$ wide wings. We were not
able to obtain echelle measurements of \hbox{[Si\,{\sc ix}]}
3.934$\mu$m, the highest excitation line so far detected from
NGC~6302. Transmission modulation by the fringes completely dominates
the emission from \hbox{[Si\,{\sc ix}]}, which is more than a thousand
times fainter than the \hbox{[Ar\,{\sc vi}]} line.

A resolving power of $\sim$20\,000 can be achieved with the echelle
and UKIRT's CGS4 spectrometer, corresponding to a velocity resolution
of 15 km~s$^{-1}$. With the short camera and a one pixel wide slit,
the pixel size is about 1.5$''$ spatially and 1$''$ spectrally. We
oversampled the spectra by shifting the array 3 times over 2 pixels.

\begin{figure*}
\begin{center}
\resizebox{\textwidth}{!}{\epsfig{file=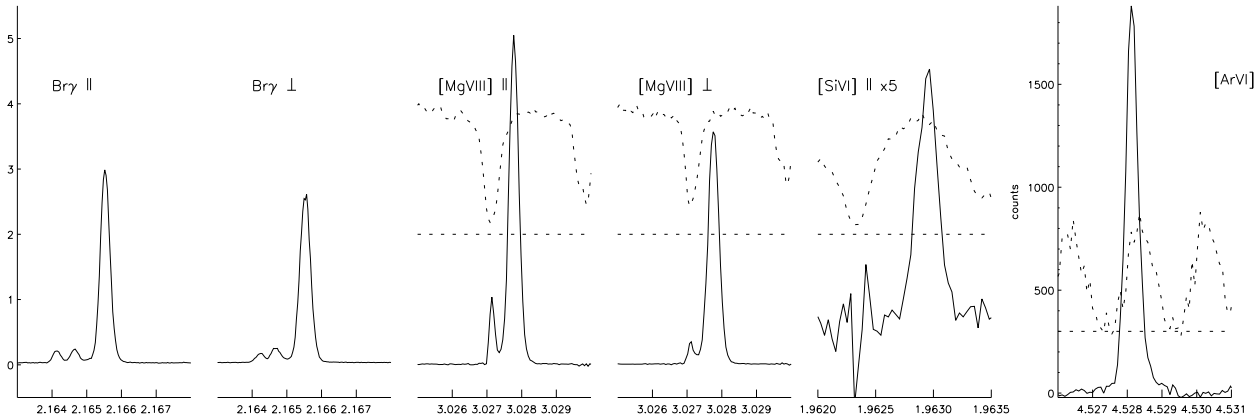}}
\end{center}
\caption{Lines observed from NGC~6302 with the CGS4 echelle. Flux
densities are given in 10$^{-12}$~W\,m$^{-2}$\,${\mu}$m, and the
calibrated wavelengths for the nebula's rest frame are in microns. The
dotted lines show the standard star spectra used to flux calibrate the
observations, doppler shifted to the rest frame of NGC~6302. The
feature in the short wavelength wing of the [Mg~{\sc viii}] line could
not be identified. The dotted curve over the [Ar~{\sc vi}] spectrum
corresponds to the CVF fringe pattern. The rms noise level is
$5\times10^{-15}$~W\,m$^{-2}$\,${\mu}$m in the [Mg\,{\sc viii}]
spectra, and 19 counts in the [Ar\,{\sc vi}] spectrum.}
\label{fig:echelle_n6302}
\end{figure*}

\begin{figure}
\begin{center}
\resizebox{8cm}{!}{\epsfig{file=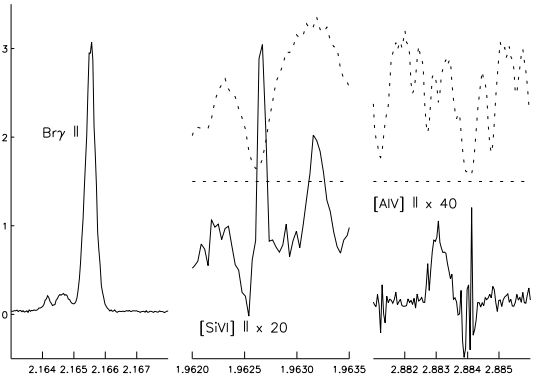}}
\end{center}
\caption{ Lines observed from NGC~6537. Flux
densities are given in $10^{-12}$~W~m$^{-2}~{\mu}$m, and wavelengths are
in microns and in the nebula's rest frame. The dotted lines are the
standard star spectra used to flux calibrate the observations,
doppler shifted to NGC~6537's rest frame.}
\label{fig:echelle_n6537}
\end{figure}

Figure \ref{fig:echelle_n6302} and \ref{fig:echelle_n6537} show the
spectra obtained by collapsing all the emission along the slit.  In
the cases where it was relevant (i.e. when there was significant
structure), the spectrum of a standard star is superposed in dotted
lines on the nebular ion's spectrum.  The standard star spectra are shown
in arbitrary units, but zero counts is represented by the horizontal
dotted lines. Figures \ref{fig:echelle} and \ref{fig:frames_n6537}
contain the reduced frames for the lines observed with CGS4+echelle.

The secondary line at 3.0271$\mu$m apparent in the \hbox{[Mg\,{\sc
viii}]} spectra for both slit positions could not be identified.  It
is not believed to be a \hbox{[Mg\,{\sc viii}]} 3.028$\mu$m velocity
component at $-64$~km\,s$^{-1}$ because it is not seen in the spectra
of the other ions. Neither is it an atmospheric transmission feature,
which can be concluded by inspecting the frames presented in
Figure~\ref{fig:echelle}, nor is it a leak from neighbouring
orders. It is nonetheless probable that the secondary line originates
from a coronal ion formed with an ionization potential between that
required to produce \hbox{Si\,{\sc vi}} and \hbox{Si\,{\sc ix}}
(i.e. from 166.77 eV to 303.16 eV), since the low resolution spectrum
of NGC~6537 shows no sign of any line at all near 3.028$\mu$m. A
Doppler shift of the secondary line deeper into the atmospheric
transmission trough seen in Figure \ref{fig:echelle_n6302} is
discarded on the basis that at the time of the echelle observations,
the velocity of NGC~6302 was $-49\pm$1~km\,s$^{-1}$, in the Earth's
rest frame, whereas at the time of the low resolution observations,
NGC~6537 was at $-42\pm$3~km\,s$^{-1}$ (see Appendix \ref{sec:acc}).


In the Br$\gamma$ frame (Figure~\ref{fig:echelle_n6302}), the fluxes
and wavelengths of the two secondary lines correspond to
\hbox{He\,{\sc i}} 4-7 at 2.16415$\mu$m (the singlet
$^{1}$F$^{\mathrm{o}}$-$^{1}$G and triplet
$^{3}$F$^{\mathrm{o}}$-$^{3}$G transitions are blended) and
\hbox{He\,{\sc ii}} 8-14 at 2.16464$\mu$m, using the emissivity Tables
from Smits et al. (1991) and Hummer \& Storey (1987).


Only the raw observation is shown for \hbox{[Ar\,{\sc vi}]}. This
frame was affected by the CVF fringes, which can be seen in the
standard star's spectrum shown by the dotted line.  However a flux
calibration is possible and the total \hbox{[Ar\,{\sc vi}]} flux
through the one-pixel 1$''$ wide slit is about 10 times that for
\hbox{[Mg\,{\sc viii}]}. A higher \hbox{[Ar\,{\sc vi}]} flux relative
to \hbox{[Mg\,{\sc viii}]} is found in the low resolution spectrum,
with a 3$''$ slit, which is an indication that the \hbox{[Ar\,{\sc
vi}]} emission has a larger spatial extent than \hbox{[Mg\,{\sc
viii}]}.

\section{Abundance study and photoionization models}
\label{sec:abundances}

\subsection{Extinction estimates and emission measures}

We shall adopt for NGC~6302 the density and temperature estimated by
Oliva et al. (1996) from a variety of transitions, $T_{e}$=19\,000\,K
(e.g. from [Ar\,{\sc v}]~$\lambda$4626,$\lambda$7006 and [Ne\,{\sc
iv}]~$\lambda$2422,$\lambda$4725), and $N_\mathrm{e}$=18000cm$^{-3}$. For
NGC~6537 we used the [O\,{\sc iii}] electron temperature
$T_{e}$=16000\,K and $N_\mathrm{e}$=17000cm$^{-3}$ from Feibelman et
al. (1985). In the case of a photoionization dominated structure the
temperature profile across a uniform nebula is expected to be fairly
flat around $T_{e}$=20\,000~K ({\small CLOUDY} models at constant proton
density, see below).  We will also argue in Section \ref{sec:kine}, in
a comparison between the spatio-kinematical data and the structure of
model nebulae, that the physical conditions must be quite uniform
across the cores of the nebulae, at least when averaged over 1$''$
scales.

We used the relative intensities between H\,{\sc i} recombination
lines to evaluate errors in flux calibration. The errors stem from
variations in slit positions for different frames, approximate
estimations of standard star fluxes, and unresolved atmospheric
transmission features.  A restricted range in the J window including
Pa$\beta$ at 1.282$\mu$m and the 3$^{3}$S-4$^{3}$P \hbox{He\,{\sc i}}
transition at 1.2792 was also observed in both nebulae with CGS4 in
grating mode. Atmospheric transmission is quite good for Pa$\beta$,
and it is a bright line observed at very good signal to
noise. Pa$\beta$ was therefore used as the flux reference H\,{\sc i}
line. On the other hand, Pa$\beta$ is appreciably more affected by
extinction than the longer wavelength IR lines, but its dereddened
flux can be estimated from free-free radio continuum maps. We used the
relation given by Milne \& Aller (1975) to estimate the total H$\beta$
flux from the 5 GHz free-free continuum measurement. We follow
similar  assumptions as in AH88, that the He abundance
relative to H\,{\sc i} is 18 per cent, and the fraction of the He
atoms that are doubly ionized is 75 per cent. The total Pa$\beta$ flux
can then be estimated from the emissivity tables given by Hummer \&
Storey (1987), assuming Case B. For the hydrogen lines with
\hbox{He\,{\sc i}} blending we used the \hbox{He\,{\sc i}}
emissivities of Smits (1991), normalised to \hbox{He\,{\sc i}}
3$^{3}$S-4$^{3}$P.


Gomez et al. (1989) present a VLA map of NGC~6302 at 5~GHz, which
allows extinction estimates for a 3$\times$3 arcsec$^{2}$ beam. Our
measured flux for Pa$\beta$ of 3.9$\times$10$^{-15}$~W\,m$^{-2}$ gives
$A_\mathrm{V}=3.7$ in the central region, with $S$(5\,GHz)=375\,mJy
(over the same region we have 1.6$\times$10$^{-16}$~W\,m$^{-2}$ for
the \hbox{He\,{\sc i}} 3$^{3}$S-4$^{3}$P observed flux). Although this
value is not in agreement with the mean value for the whole nebula of
$A_\mathrm{V}=2.65$ (Pottasch et al. 1996), it agrees with the
conclusion by AH88 that extinction in the central regions of NGC~6302
is higher than the mean by 1.3 magnitudes.  Unfortunately no VLA maps
are available for NGC~6537, but the estimations of AH88 give
$A_\mathrm{V}=4$, very close to the value for NGC~6302 (for which they
estimate $A_\mathrm{V}=4.3$). We fixed the extinction in NGC~6537 to
$A_\mathrm{V}=3.5$ magnitudes by requiring the unblended Br$\alpha$ to
Pa$\beta$ flux ratio to match the ratio of their emissivities at
20\,000~K (from Hummer \& Storey 1987).

Unresolved atmospheric transmission (AT) troughs can be recognised from an
atmospheric transmission model for Mauna Kea (based on the HITRAN
database, Rothman et al. 1992). We took Pf\,5-10 as typical, since it
lies in a trough, but its predicted flux was very close to the
observed one. Unresolved atmospheric transmission features are however
a source of uncertainty in the low resolution flux measurements, as
shown by AH88 in the case of \hbox{[Si\,{\sc vi}]}, which can be
completely absorbed by atmospheric H$_2$O. We used the extinction law
given by Rieke \& Lebofsky (1985), and the dereddened and deblended
H\,{\sc i} fluxes are at most 20 per cent different from the values
expected from the radio map in the case of NGC~6302. Observing from
Mauna Kea, with a low column of precipitable water vapour, minimizes
the effect of AT troughs. 

The emission measure integrated along the line of sight, $EM=\int\, ds~
N_\mathrm{p}\,N_\mathrm{e}$, can be estimated from the intensity of hydrogen emission
lines (e.g. Osterbrock 1989),
\begin{equation}
I= \int ds~ N_\mathrm{p}N_\mathrm{e} (h\nu ~ \alpha_\mathrm{eff} )/(4 \pi),
\end{equation}
where $\alpha_\mathrm{eff}$ is the effective recombination coefficient
for Pa$\beta$ (Hummer and Storey 1987). The Pa$\beta$ line flux from
the central 3$''$x3$''$ in NGC~6302 is
$F_{\mathrm{Pa}\beta}$=3.9$\times$10$^{-15}$~W\,m$^{-2}$, giving
$EM=5.6{\times}10^{35}$~m$^{-5}$. For NGC~6537,
$F_{\mathrm{Pa}\beta}$=3.8$\times$10$^{-15}$~W\,m$^{-2}$ , and
$EM=5{\times}10^{35}$~m$^{-5}$.

\subsection{Abundance patterns and photoionization models}
\label{subsection:photo}

Table \ref{table:abundances} lists the deblended and dereddened
surface brightnesses through the central CGS4 row ($3''\times3''$
square aperture), together with the abundances calculated using the
atomic data referenced there. The [Si\,{\sc vi}] and [Si\,{\sc vii}]
intensities were derived from UKIRT observations on April 1986 with
the CVF on UKT9, using a 19.6 arcsec circular aperture. We also used
some fluxes from Pottasch et al. (1996) and Rowlands et al. (1994), as
indicated in Table \ref{table:abundances}. We estimated the average
intensity in the central $3''\times3''$ by using the spatial
information from our grating and echelle CGS4 spectra (see the end of
Appendix \ref{sec:acc}). The fluxes obtained with large beams were
scaled by a factor 0.4 for NGC~6302 and 0.58 for NGC~6537, to estimate
the corresponding fluxes in the central $3''\times3''$ (except for
\hbox{[S\,{\sc iii}]}, for which we used a factor of 0.15 for
NGC~6302, and 0.37 for NGC~6537).

Although we list only ionic abundances, we can draw some conclusions
about abundance patterns from those elements having
several observed ionization stages and  reliable line fluxes. 
Magnesium seems to be
slightly depleted relative to solar, but not as strongly as
aluminium. Potassium shows a very small ionization fraction in K\,{\sc
vii}, when compared to \hbox{Ne\,{\sc v}} at a similar ionization
potential, or to Na\,{\sc vi} and \hbox{Na\,{\sc vii}}. Although a
somewhat higher abundance in neon is expected in Type I PNe (Corradi
\& Schwarz 1993), it could be that potassium is slightly depleted,
since no special overabundance is expected in sodium.

\begin{table*}
\begin{center}
\caption{Ionic abundances by number for NGC~6302 and NGC~6537, relative to solar values.} \label{table:abundances}
\begin{tabular}{ll|rllllcr} \hline
                   &                                    &                  &    \multicolumn{2}{c}{NGC~6302} &  \multicolumn{2}{c}{NGC~6537}  \\       
                   &                                    &$\lambda$[$\mu$m] &      I$_\mathrm{dered}$ &    abundance~$^d$     &      I$_\mathrm{dered}$ &    abundance~$^d$    &         IP&  atomic \\      
           	   &                                    &                  & W\,m$^{-2}$\,sr$^{-1}$  &                       & W\,m$^{-2}$\,sr$^{-1}$  &                      &  eV &  data~$^c$  \\ \hline 
       H\,{\sc i}  &          3-5                       &   1.282&   4.15$\times10^{-}$                &   ---                 &    4.73$\times10^{-5}$       &   ---                &     13.6  & 1 \\   
       He\,{\sc i} &          3$^{3}$S-4$^{3}$P         &   1.253&    5.89$\times10^{-7}$              &    0.47               &    3.23$\times10^{-7}$       &    2.68              &     24.6  & 2  \\        
      He\,{\sc ii} &                6-7                 &  3.0917&    1.49$\times10^{-5}$              &    1.38               &    1.62$\times10^{-5}$       &    1.60              &     54.4  & 1 \\         
      Ne\,{\sc ii} &   $^{2}$P$_{3/2}$-$^{2}$P$_{1/2}$  &   12.81&   1.21$\times10^{-4}$               &  0.36                 &   4.24$\times10^{-5}$        &  0.17                &     21.6  & 3   \\      
      Ne\,{\sc iii}&   $^{3}$P$_{2}$-$^{3}$P$_{1}$      &   15.5 &   7.73$\times10^{-4}~^{a^\prime}$   &  1.2                  &      ---                     & ---                  &     41.0  & 4    \\     
      Ne\,{\sc iii}&   $^{3}$P$_{1}$-$^{3}$P$_{0}$      &   36.0 &   4.27$\times10^{-5}~^{a^\prime}$   &  0.9                  &      ---                     & ---                  &     41.0  & 4   \\      
      Ne\,{\sc v}  &       $^{3}$P$_{1}$-$^{3}$P$_{2}$  &   14.32&   1.42$\times10^{-3}~^a$            &  0.36                 &   4.75$\times10^{-4}~^{b}$   &  0.13                &     97.1  & 5   \\      
      Ne\,{\sc v}  &       $^{3}$P$_{0}$-$^{3}$P$_{1}$  &   24.31&   5.31$\times10^{-4}~^a$            &   0.37                &   2.60$\times10^{-4}~^{b}$   &   0.19               &     97.1  & 5   \\      
      Ne\,{\sc vi} &   $^{2}$P$_{1/2}$-$^{2}$P$_{3/2}$  &   7.64 &    1.49$\times10^{-3}$              &   0.18                &   4.89$\times10^{-4}$        &   0.12               &     126   & 6   \\      
      Na\,{\sc vi} &       $^{3}$P$_{1}$-$^{3}$P$_{2}$  &   8.64 &   2.20$\times10^{-5}$               &   0.33                &   8.95$\times10^{-6}$        &   0.15               &     138   & 5   \\      
     Na\,{\sc vii} &   $^{2}$P$_{1/2}$-$^{2}$P$_{3/2}$  &  4.690 &   1.89$\times10^{-5}$               &   0.12                &      ---                     & ---                  &     172   & 7   \\      
       Mg\,{\sc v} &       $^{3}$P$_{2}$-$^{3}$P$_{1}$  &  5.608 &   7.99$\times10^{-5}~^a$            &  0.11                 &      ---                     & ---                  &     109   & 4  \\       
     Mg\,{\sc vii} &       $^{3}$P$_{1}$-$^{3}$P$_{2}$  &  5.502 &   6.15$\times10^{-5}~^a$            &   4.9$\times10^{-2}$  &      ---                     & ---    		     &     187   & 5   \\      
    Mg\,{\sc viii} &    $^{2}$P$_{1/2}$-$^{2}$P$_{3/2}$ &  3.028 &    8.55$\times10^{-6}$              &    2.8$\times10^{-3}$ &      ---                     & ---    		     &     225   & 6  \\       
       Al\,{\sc v} &    $^{2}$P$_{3/2}$-$^{2}$P$_{1/2}$ &  2.883 &    5.86$\times10^{-8}$              &    1.0$\times10^{-3}$ &      ---                     & ---    		     &     120   & 3 \\        
      Al\,{\sc vi} &        $^{3}$P$_{2}$-$^{3}$P$_{1}$ &  3.659 &    6.66$\times10^{-7}$              &    1.9$\times10^{-3}$ &      ---                     & ---                  &      154  & 4 \\        
      Si\,{\sc vi} &   $^{2}$P$_{3/2}$-$^{2}$P$_{1/2}$  &  1.96  &   1.11$\times10^{-4}$               &   0.22                &   5.88$\times10^{-5}$        &   0.13               &     167   & 3  \\       
     Si\,{\sc vii} &       $^{3}$P$_{2}$-$^{3}$P$_{1}$  &  2.48  &   9.70$\times10^{-5}$               &   0.14                &   1.59$\times10^{-6}$        &   2.6$\times10^{-3}$ &     205   & 4  \\       
      Si\,{\sc ix} &        $^{3}$P$_{0}$-$^{3}$P$_{1}$ &  3.934 &    1.42$\times10^{-7}$              &    3.0$\times10^{-5}$ &    ---                       & --- 	             &     303   & 5  \\       
      S\,{\sc iii} &       $^{3}$P$_{1}$-$^{3}$P$_{2}$  &  18.71 &   4.17$\times10^{-5}~^{b}$          &   0.11                &   7.68$\times10^{-5}~^{b}$   &   0.22  	     &    23.3   & 8    \\     
       S\,{\sc iv} &   $^{2}$P$_{1/2}$-$^{2}$P$_{3/2}$  &  10.50 &   2.08$\times10^{-4}$               &   0.13                &   2.91$\times10^{-4}$        &   0.38               &    34.8   & 9    \\     
     Ar\,{\sc iii} &       $^{3}$P$_{2}$-$^{3}$P$_{1}$  &   9.01 &   7.66$\times10^{-5}$               &   0.38                &   3.44$\times10^{-5}$        &   0.19               &    27.6   & 8    \\     
      Ar\,{\sc v}  &       $^{3}$P$_{1}$-$^{3}$P$_{2}$  &   7.91 &   4.29$\times10^{-5}$               &   8.4$\times10^{-2}$  &   2.91$\times10^{-5}$        &   6.2$\times10^{-2}$ &    59.8   & 8    \\     
      Ar\,{\sc v}  &       $^{3}$P$_{0}$-$^{3}$P$_{1}$  &  13.11 &   2.09$\times10^{-5}$               &   5.1$\times10^{-2}$  &   2.22$\times10^{-5}$        &   5.8$\times10^{-2}$ &    59.8   & 8    \\     
      Ar\,{\sc vi} &    $^{2}$P$_{1/2}$-$^{2}$P$_{3/2}$ &  4.528 &    1.37$\times10^{-4}$              &   0.11                &    3.04$\times10^{-4}$       &   0.32               &    75.0   & 10 \\       
      K\,{\sc iii} &    $^{2}$P$_{3/2}$-$^{2}$P$_{1/2}$ &  4.616 &    1.52$\times10^{-6}$              &   0.20                &    5.09$\times10^{-7}$       &   7.3$\times10^{-2}$ &    31.6   & 11  \\      
      K\,{\sc vii} &    $^{2}$P$_{1/2}$-$^{2}$P$_{3/2}$ &  3.189 &    1.32$\times10^{-6}$              &    3.3$\times10^{-2}$ &    9.83$\times10^{-7}$       &  2.7$\times10^{-2}$  &    99.9   & 10 \\ \hline
\end{tabular}																					  
																						      
\medskip
\end{center}

$^{a}$ {\it ISO} measurement,  Pottasch et al. (1996);  $^{a^\prime}$ Pottasch \& Beintema (1997)\\
$^{b}$ Rowlands et al. (1994).\\
$^{c}$ The collision strengths were for the most part taken from the
IRON project.  1, Hummer \& Storey (1987); 2, Smits (1991); 3,
Saraph \& Tully 1994; 4, Butler \& Zeipen (1994); 5, Lennon \&Burke
(1994); 6, Zhang (1994); 7, Ferland (1996, as used in {\small
CLOUDY}); 8, Galav\'{\i}s et al. (1995); 9, Johnson et al. (1986); 10,
Saraph \& Storey (1996); 11, Pelan \& Berrington (1995) \\
$^{d}$ Ionic abundance relative to the total solar abundance of each
element, or (Nx/Np)/solar. Solar abundances from Grevesse and Anders
(1989), Grevesse and Noels (1993).\\
\end{table*}

We also used {\small CLOUDY} 90.03 (Ferland 1996) to construct a
photoionization model of NGC~6302.  The nebula is simulated as a
spherically symmetric shell, although the fluxes of emission lines are
calculated by truncating the shell in order to simulate the long slit
observations. The central star's radiation field is assumed to be that
of a blackbody. No systematic depletion of metals on grains was
assumed, principally because little is known about dust survival in
PNe. Nonetheless, the abundances of elements likely to be depleted
were allowed to vary, in particular magnesium shows an important line
that should be reproduced accurately. The abundances were otherwise
taken as solar. We also varied the filling factor, the blackbody's
temperature and the inner radius of the nebula. {\small CLOUDY}'s
optimisation driver was constrained by the fluxes of IR
coronal ions and the H\,{\sc ii} column density (estimated from the
observed emission measure, with $N_\mathrm{e} =$ 1.8$\times10^{4}$~cm$^{-3}$).

The main problem in using {\small CLOUDY} for this application is that
we are probing the central region of NGC~6302, whereas {\small CLOUDY}
calculates the fluxes from the whole nebula. The cumulative fluxes
versus radius could be match\-ed with our spectra, leaving the cutoff
radius as yet another free parameter. But one way around this problem
is to reproduce the fluxes relative to another ionic line. We took
\hbox{[Ar\,{\sc vi}]} at 4.53~$\mu$m as the reference line, and the
{\small CLOUDY} output can be seen in Table \ref{c90_output}, as well
as the best fit values for the free parameters. A total hydrogen volume
density of $\log(N_{\mathrm{H}})$=4.2 was given as input, together
with a relative helium abundance by number of 0.14 (using 0.18, as
reported by Aller et al. 1981 and as derived in this work, does not change
the results).  The total luminosity of the central star was fixed to be
8500\,L$_{\odot}$, simply because it is close to a standard PN value,
although Pottasch et al. (1996) estimate $L=11000\,$L$_\odot$ for a
distance of 1.6\,kpc (Gomez et al. 1993). Using 11000\,$L_\odot$
leaves the model unchanged. The calculation stopped at an outer radius
of 1.2$\times10^{17}$\,cm because the package's lower limit electron
temperature of 4$\times10^{3}$\,K was reached.

\begin{table}
\caption{{\footnotesize CLOUDY} output from the best fit models for
NGC~6302 and NGC~6537. Columns 2--5 list ionic line fluxes
relative to \hbox{[Ar\,{\sc vi}]}~4.53$\mu$m~=~1.00.}
\label{c90_output}
\small
\begin{tabular}{lll}\hline
               & NGC~6302                 & NGC~6537    \\
 filling factor&	 0.70             & 0.58                    \\
 inner radius	&  6.2$\times10^{15}$~cm  & 3.8$\times10^{15}$~cm        \\
 blackbody T$_{\star}$&		 241\,000~K    & 156\,000~K    \\
 luminosity   &          8500~L$_{\odot}$     & 8500~L$_{\odot}$  \\
 Mg abundance & 	$\approx$ 0.5 solar   &  ---     \\
 Al abundance & 	$\approx$ 0.01  solar & ---      \\
\end{tabular}
\begin{tabular}{l|ccccc}\hline  
                          & \multicolumn{2}{c}{NGC~6302} &  \multicolumn{2}{c}{NGC~6537}  \\
 ID~~~$\lambda$($\mu$m)   &       Model  &  Obs.   &     Model  &  Obs. \\ \hline
$[$Ne\,{\sc v}$]$   14.3&     5.7&     10  &      3.6&     1.6             \\
$[$Ne\,{\sc v}$]$   24.2&     1.8&     3.9 &     1.2&     0.86\\
$[$Ne\,{\sc vi}$]$  7.64&     8.6&     11  &    1.3&     1.6\\
$[$Na\,{\sc vii}$]$   4.69&    7.6\,$10^{-2}$ &     0.14   & --- & ---\\
$[$Mg\,{\sc v}$]$   5.60&     0.30&     0.60   & --- & ---\\
$[$Mg\,{\sc vii}$]$  5.51&     0.31&     0.40  & --- & ---\\
$[$Mg\,{\sc viii}$]$   3.03&   5.0\,$10^{-2}$  &     6.0\,$10^{-2}$  & --- & ---\\
$[$Al\,{\sc v}$]$   2.9&      6.0\,$10^{-4}$&     5.0\,$10^{-4}$  & --- & ---\\
$[$Al\,{\sc vi}$]$   3.66&     4.8\,$10^{-3}$&     5.0\,$10^{-3}$\\
$[$Si\,{\sc vi}$]$   1.96&     0.80&     0.80 &     9.0\,$10^{-2}$ &  0.19\\
$[$Si\,{\sc vii}$]$   2.47&     0.32&     0.71  &    2.5\,$10^{-3}$& 5.2\,$10^{-3}$\\
$[$Si\,{\sc ix}$]$   3.93&     1.2\,$10^{-3}$&     1.0\,$10^{-3}$  & --- & ---\\
$[$S \,{\sc iv}$]$   10.5&     2.26&     1.52 &     2.89&     0.96\\
$[$Ar\,{\sc v}$]$   7.91&     9.3\,$10^{-2}$&     0.30 &     0.12 &   9.6\,$10^{-2}$\\
$[$Ar\,{\sc v}$]$   13.1&     7.6\,$10^{-2}$&     0.15 & 9.5\,$10^{-2}$&  7.3\,$10^{-2}$\\
$[$Ar\,{\sc vi}$]$  4.53&     1.00 &     1.00 & 1.00 &  1.00 \\
$[$Na\,{\sc vi}$]$   8.64&     8.3\,$10^{-2}$&     0.16  & 1.7\,$10^{-2}$&   2.9\,$10^{-2}$ \\
$[$K \,{\sc vii}$]$   3.19&     5.2\,$10^{-2}$&     1.0\,$10^{-2}$  &  2.5\,$10^{-2}$&     3.2\,$10^{-3}$\\
  ~                  &         ~             &      ~ \\
 N(H\,{\sc ii})$^{a}$      &  1.49\,$10^{21}$&  1.55\,$10^{21}$  &   1.33\,$10^{21}$&  1.40\,$10^{21}$ \\ \hline
\end{tabular}
\medskip

$^{a}$ Column density of H$^{+}$ in cm$^{-2}$ integrated outwards from
the central star.
\end{table}

The best fit black body temperature is about 250\,000~K. This is
to be compared to the very high values of 450\,000~K found by Lame \&
Ferland (1991), and 380\,000~K by Pottasch et al. (1996). One
important new datum used here is the \hbox{[Si\,{\sc ix}]} flux, which is
actually quite low and would be significantly
overestimated by models with $T_{\star}$ higher than 250\,000~K. This
result has already been mentioned by Marconi et al. (1996) in their
study of the Seyfert 2 galaxy NGC~1068. Arguing against the production
of \hbox{Si\,{\sc ix}} by hot stars, they take as an example the
nucleus of NGC~6302, and model the absence of \hbox{[Si\,{\sc ix}]}
from their NGC~6302 spectra by putting an upper limit on $T_{\star}$
of 250\,000~K (using {\small CLOUDY} as well). The central star in
NGC~6302 is thus probably not as hot as previously thought,
$T_{\star}$=380\,000~K overestimates the \hbox{[Si\,{\sc ix}]} flux by
a factor 60 (with an optimised filling factor of 0.16, and an inner
radius of 3.1$\times10^{16}$ cm). But could silicon be depleted?  The
{\small CLOUDY} model shows that 70\% of all silicon nuclei are
distributed in Si\,{\sc v}, Si\,{\sc vi} and Si\,{\sc vii}. The match
to the [Si\,{\sc vi}] and [Si\,{\sc vii}] lines should provide a good
grip on the silicon abundance, and they are approximately reproduced
using a solar silicon abundance.


Together with the model results for NGC~6302, Table \ref{c90_output}
shows the result of a CLOUDY model that reproduces the lines observed
from NGC~6537, with a central star temperature of 156\,000~K (using
similar inputs as for NGC~6302, with $\log(N_\mathrm{H})=4.2$; the
calculation stopped because the package's lower limit electron temperature
was reached).  [Al\,{\sc vi}] was not detected towards NGC~6537, but its
predicted flux assuming a solar Al abundance would be
5.7$\times10^{-2}$ times the [Ar\,{\sc vi}] flux (which is
6.4\,$\times10^{-14}$~W\,m$^{-2}$ for the central 3$''\times3''$ in
NGC~6537). As the [Al\,{\sc vi}] frame shows Hu\,17 close to the
detection limit, with a flux 5.4$\times10^{-4}$ times the [Ar\,{\sc vi}]
flux, we can conclude than Al is depleted to be less than one hundredth
solar.

\begin{figure}
\begin{center}
\resizebox{8cm}{!}{\epsfig{file=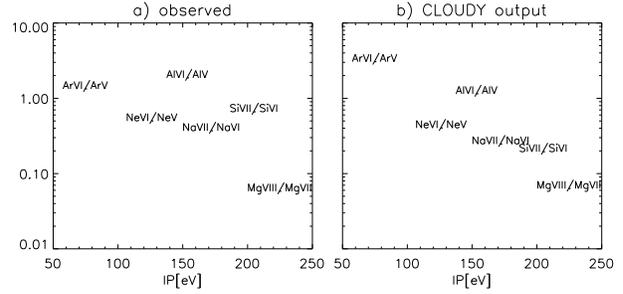}}
\end{center}
\caption{ Ionization distribution for NGC~6302. The abscissa is the
photon
energy required to ionize the lower stage of ionization, and the
ordinate is ${\mathrm{N(X^{+i+1})/N(X^{+i})}}$. The uncertainty in the
observed ratios can be up to a factor of 2.}
\label{fig:icurve}
\end{figure}

The abundance ratios of consecutive stages of ionization as a function
of ionization potential are, in the case of photoionization, a probe of
the high energy flux from the central star.  Figure~\ref{fig:icurve}a
shows the ionization distribution calculated from the observed fluxes.
We can use the {\small CLOUDY} model for NGC~6302 to construct a similar
curve, shown in Figure~\ref{fig:icurve}b. The curve is in remarkable
agreement with the observations, except for the silicon ions. The
Mg\,{\sc viii} to Mg\,{\sc vii} ratio also argues for a $T_{\star}$ of
about 250\,000~K. Such good agreement is surprising considering all
the approximations involved in the modelling. In a complete model of
the nebula, a number of arbitrary assumptions about the recombination
data have to be made from the very first ionization stage. The
abundance ratio of consecutive stages of ionization is inversely
proportional to the total recombination coefficient, and little
information is available on the dielectronic part, especially at
nebular temperatures. But even if the atomic parameters were known
with sufficient accuracy, as well as the temperature of the coronal
line region, the central star's spectrum may differ significantly from
a blackbody.

With the cautions noted above (that a match between model and observed
fluxes is acceptable within a factor of $\sim$2), the abundances
computed by {\small CLOUDY} confirm the patterns from the direct
estimate. For both NGC~6302 and NGC~6537, aluminium is depleted to one
hundredth of the solar value, or even less. The magnesium abundance is
about half solar in NGC~6302. Neon seems to be overabundant in NGC~6302,
but by
less than a factor of two relative to the solar value. It could be
that potassium is depleted to one fifth of the solar abundance in
NGC~6302 as well as in NGC~6537, although only one line is
available. The uncertainties in the photoionization model may be
reflected in the discrepant \hbox{Si\,{\sc vii}} to \hbox{Si\,{\sc
vi}} abundance ratio (which is sensitive to extinction) and the model
deficiency in [ArV] and in the sodium lines, but in all cases the
discrepancies are less than a factor of 2--3.


Aller et al. (1981) found that Ca was depleted to one tenth solar in
NGC~6302, while Oliva et al. (1996) estimated Ca to be depleted by a
factor $\sim$20, and Fe by a factor of $\sim10^{3}$ (see also Shields
1983 for depletion patterns in other PNe). It could be that titanium
is also strongly depleted. According to the {\small CLOUDY} model, the
flux of [Ti\,{\sc vi}]\,1.7$\mu$m should be a little less than that of
Br$\gamma$, at about one fiftieth of \hbox{[Ar\,{\sc vi}]}, but no
spectrum in the H window has been published for NGC~6302. However,
most of the titanium should be in Ti\,{\sc vii} or Ti\,{\sc viii}, and
the ground state, magnetic dipole, [Ti\,{\sc vii}]~4144.2\,\AA~ line
is missing in the spectrum published by Aller et al. (1981).  So it
seems likely that Ti is strongly depleted.  The strong depletion of
Al, and possibly of Ti, implies that some dust is well mixed with the
gas phase, in the form of Ti and Al compounds, such as perovskite,
CaTiO$_{3}$, and corundum, Al$_{2}$O$_{3}$, which have very high
sublimation temperatures. Gail and Sedlmayr (1998) argue that the
first condensates in circumstellar outflows are Ti compounds, which
could be nucleation sites for the aluminium compounds and silicates.
When the AGB ejecta are heated to nebular temperatures and exposed to
hard UV radiation, grain ablation could explain the observed
depletions. Kingdon \& Ferland (1997 and references therein) suggest
that strong ionic depletions are indicative of dust survival in the
ionized phase of PNe. Examples are Fe depletions inferred from
Fe\,{\sc vii}, which requires radiation harder than 100~eV. We note
here that since 154\,eV is required to ionize Al\,{\sc v}, the Al
depletion derived for NGC~6302 from Al\,{\sc vi} provides evidence for
dust survival in very hard radiation field environments.  Pwa et
al. (1984, 1986) also measured Al depletions of a factor $\sim$100 for
NGC~6543 and a factor $\sim$10 for BD+30$^{\circ}$3639, from {\em IUE}
observations of Al\,{\sc ii} and Al\,{\sc iii} absorption lines.

It would seem that photoionization models can account for the observed
ionization distribution in broad terms. Zhekov \& Perinotto (1996)
attempted to model the IR coronal line emission from high excitation
PNe for the case of a collisionally excited plasma. They used an
interacting wind model, following Weaver et al. (1977), in which the
slow AGB wind is accelerated into a shock by the fast white dwarf
wind. The hot bubble grows as the shock expands and `vaporizes'
material from the cold shell. After working out the temperature
structure of the bubble, they calculated its coronal line emission
using the hot plasma emissivities. But their predicted fluxes are in
strong disagreement with those observed from NGC~6302, e.g. the
predicted [Si\,{\sc ix}] and [Mg\,{\sc viii}] fluxes should be of the
same order according to their model.

\subsection{NGC 6302's progenitor}

In this Section we present arguments for NGC\,6302's progenitor 
having a lower ZAMS mass than previously thought. The total luminosity
of the central star can be estimated from a sum of all flux
measurements of NGC~6302 from the UV to the infrared (Pottasch et
al. 1996 and references therein). For a distance of 1.6~kpc (as
determined by Gomez et al. 1993 using VLA observations of the
expansion rate), the central star's luminosity is about
$1.1{\times}10^{4}$~L$_{\odot}$. We can now locate the star in the H-R
diagram and compare with the theoretical tracks of Bl\"{o}cker (1995).
A temperature of 380\,000~K would locate the star on the track of a
hydrogen burning model with a 7\,M$_{\odot}$ initial mass (and a core
mass of $M_\mathrm{H}$~=~0.940\,M$_{\odot}$) 60 years or so after
leaving the AGB. But our estimate of 250\,000\,K would fall on a track
intermediate between the Bl\"{o}cker (1995) models with
$(M_\mathrm{ZAMS}, M_\mathrm{H})=(4\,M_\odot, 0.696\,M_{\odot})$ and
$(5\,M_\odot, 0.836\,M_{\odot})$, or a 4--5~M$_{\odot}$ initial mass,
for either the hydrogen burning models or the born-again helium
burners. It is difficult to distinguish between the two burning modes,
as the time-scale is slower in the helium burners by a factor of 3
only.  If the bulk of the nebular material was ejected at the time the
star left the AGB, the age of about 2000 years necessary to account
for NGC 6302's size of $10^{17}$~cm at an expansion speed of
10~km\,s$^{-1}$ may be better explained in terms of the born-again
models of Bl\"{o}cker (1995).  The PN progenitor could have had a ZAMS
mass of $\sim$4~M$_{\odot}$ and have left the AGB on the verge of a
thermal
pulse, since the time-scale for a straight evolution from the AGB as a
hydrogen burner is only 700\,yrs (alternatively the star could have
left the AGB just after a thermal pulse, and evolved on a similar
time-scale as the born-again case, although Bl\"{o}cker does not
provide tracks for this case).

\begin{figure}
\begin{center}
\resizebox{8cm}{!}{\epsfig{file=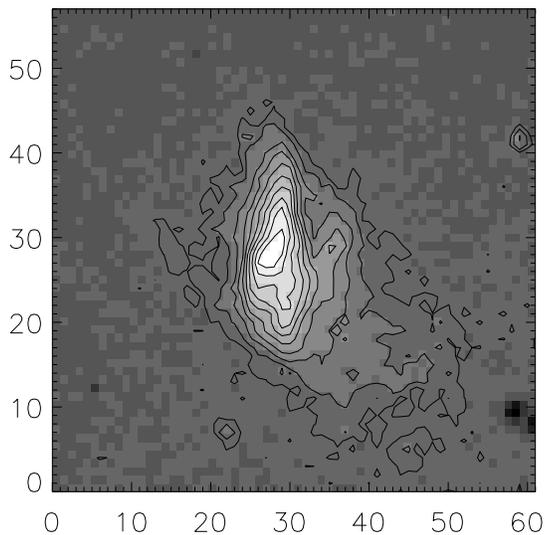}}
\end{center}
\caption{IRCAM image of the 3.3$\mu$m UIR-band emission in
NGC~6302. Pixel numbers are plotted along the axes. The pixel size
was 0.6$''$ and the width of the narrow band filter was 0.3$\mu$m. The
contour levels are .05, .1, .2, .3, .4, .5, .6, .7, .8, .9 of the peak
flux, to allow comparison with the VLA 6\,cm map from Gomez et
al. (1989). North is down and East is to the right.}\label{fig:UIRmap}
\end{figure}


The OH maser observed in NGC~6302 by Payne et al. (1988) provides evidence
for an oxygen rich environment, as does the {\em ISO} detection of
mid-IR crystalline silicate emission bands (Waters et al. 1996), and
yet the nebula shows the family of
`unidentified infrared' (UIR) emission bands, attributed to carbon
based molecules (Duley \& Williams 1981, L\'{e}ger \& Puget 1984). The
OH maser emission seems to come from the material surrounding the core
of the nebula, since the cuts in velocity published by Payne et
al. (1988) peak at different positions around the centre, over about
8$''$. On the other hand the 3.3 $\mu$m UIR emission band is centrally
peaked, as can be seen on our IRCAM frame shown in Figure
\ref{fig:UIRmap}. The extent of the UIR emission can be compared to
the VLA map of Gomez et al. (1989). The UIR bands are typical of
C-rich PNe (Cohen et al. 1986), and these observations support a
scenario in which NGC~6302's progenitor was an M giant that went
through a carbon transition at the very end of its evolution, as is
analysed in detail by Frost et al. (1998). However, maser
amplification requires particular conditions, and therefore does not
systematically trace the distribution of O-rich material, and the
evidence for a stratification in C/O chemical balance is only marginal
until high resolution OH observations are available and a model is
constructed.  Nonetheless NGC~6302 clearly includes both C- and O-rich
environments. Values of C/H = 4.4$\times10^{-4}$, N/H =
9.2$\times10^{-4}$ and O/H = 5.0$\times10^{-4}$ have been derived for
NGC 6302 by M. J. Barlow, using the UV and optical relative line
intensities measured by Barral et al.  (1982) and Aller et al. (1981),
respectively.  Collisionally excited lines were used to derive the
abundances of the singly, doubly and triply ionized species, while the
6h-5g recombination lines in the 4600~\AA\ region were used to derive
the abundances of the four-times ionized species, as in Clegg et
al. (1987).

\begin{figure}
\begin{center}
\resizebox{8cm}{!}{\epsfig{file=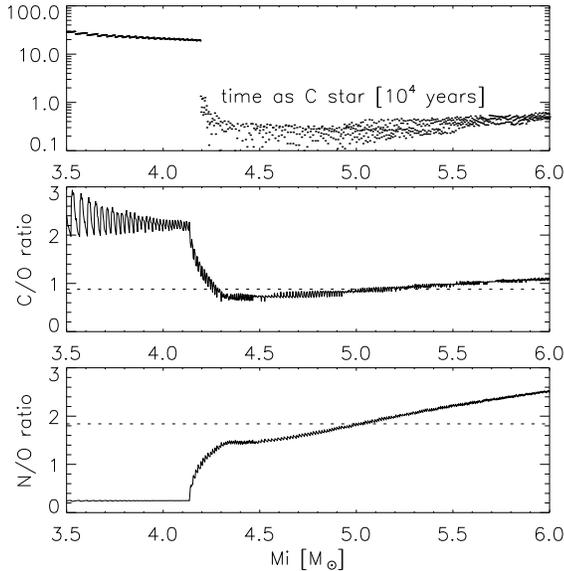}}
\end{center}
\caption{The upper plot shows the variation of the time spent
in the C star phase at the end of the AGB as a function of initial
mass. The lower two plots show the C/O and N/O ratios in the ejecta
averaged over the last 10000 years of evolution, and the horizontal
dashed lines correspond to gas phase measurements of the corresponding
ratios (see text).}\label{fig:AGB}
\end{figure}

\begin{figure}
\begin{center}
\resizebox{8cm}{!}{\epsfig{file=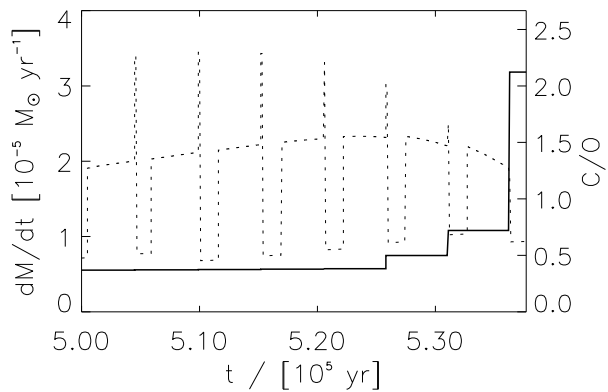}}
\end{center}
\caption{Schematic evolution of a 4.45 M$_{\odot}$ star at the end of
the AGB. The dotted line is the mass loss rate (left hand scale) and
the solid line is the surface C/O ratio (right hand scale). The last
thermal pulse changes the C/O chemical balance of the star's envelope,
approximately 2000 years before envelope ejection.}\label{fig:AGBevo}
\end{figure}

Synthetic AGB evolution models (Renzini \& Voli 1981) are compilations
of model results from full evolutionary calculations, and allow
searches over a continuous range of progenitor masses. Using an
adaptation of the synthetic AGB model by Groenewegen \& de\,Jong
(1993), we can search for a progenitor mass that accounts for the
properties of NGC~6302. In this model we do not constrain envelope
burning to stop at the same time as dredge-up (e.g. Frost et
al. 1998). Many uncertainties affect the treatment of hot bottom
burning (HBB), in this case one of them being the arbitrary onset of
HBB for core masses greater than 0.8 M$_{\odot}$\footnote{The lack of
a continuous transition would preclude the existence of type IIa PNe,
with log(N/O)$>$--0.6 (as defined by Fa\'{u}ndez-Abans \& Maciel
1987), which in Groenewegen \& de\,Jong (1993) are mainly obtained at
high initial metallicity, but are nonetheless observed to exist in the
LMC (Leisy \& Dennefeld 1996). A limiting core mass of 0.8
M$_{\odot}$ for HBB also constrains N enrichment to occur for stars
close to the 2$^\mathrm{nd}$ dredge-up critical mass (Becker \& Iben
1979), which does not reproduce the population of C-rich Type~I PNe,
except for progenitor masses greater than 6 M$_{\odot}$.} But our
purpose here is to illustrate an alternative argument to the post-AGB
tracks for NGC~6302 having a relatively low progenitor mass. The above
uncertainties in the treatment of HBB would decrease the lower mass
limit of the range of initial masses that account for NGC~6302's
composition. Figure \ref{fig:AGB} shows the time after the C
transition as a function of initial mass for stars at the end of the
AGB, as well as the averaged C/O and N/O ratios over the last 10000
years of mass loss.  Towards 6\,M$_{\odot}$ the time after the
C-transition is too long, the carbon rich nebular material having been
ejected at most over 2000 years. Below 4.2\,M$_{\odot}$ HBB is not
activated and the envelope composition is dominated by the dredge-up
of carbon. Thus, in terms of the quenching of HBB at the end of the
AGB evolution of the progenitor, the main sequence mass of NGC~6302's
progenitor would lie between 4\,M$_{\odot}$ and 5\,M$_{\odot}$ to
account for the transition in C/O ratio. This mass range is also
confirmed by the observed gas phase abundances, and is in independent
agreement with the comparison of the central star's temperature and
the post-AGB tracks of Bl\"{o}cker (1995). Figure \ref{fig:AGBevo}
shows the synthetic evolution of a 4.45 M$_{\odot}$ model that matches
the observational requirements. It can be seen that the star goes
through a C transition over the last 2000 years. The final C/O ratio
is as high as that found in circumstellar environments which show the
UIR bands (and not SiC, e.g. Barlow 1983, Roche 1989).


\section{Kinematical structure of the cores of NGC~6302 and NGC~6537}
\label{sec:kine}

\subsection{Velocity profiles from the echelle spectra and the observed
ionization stage stratification}

\begin{figure*}
\begin{center}
\resizebox{16cm}{!}{\epsfig{file=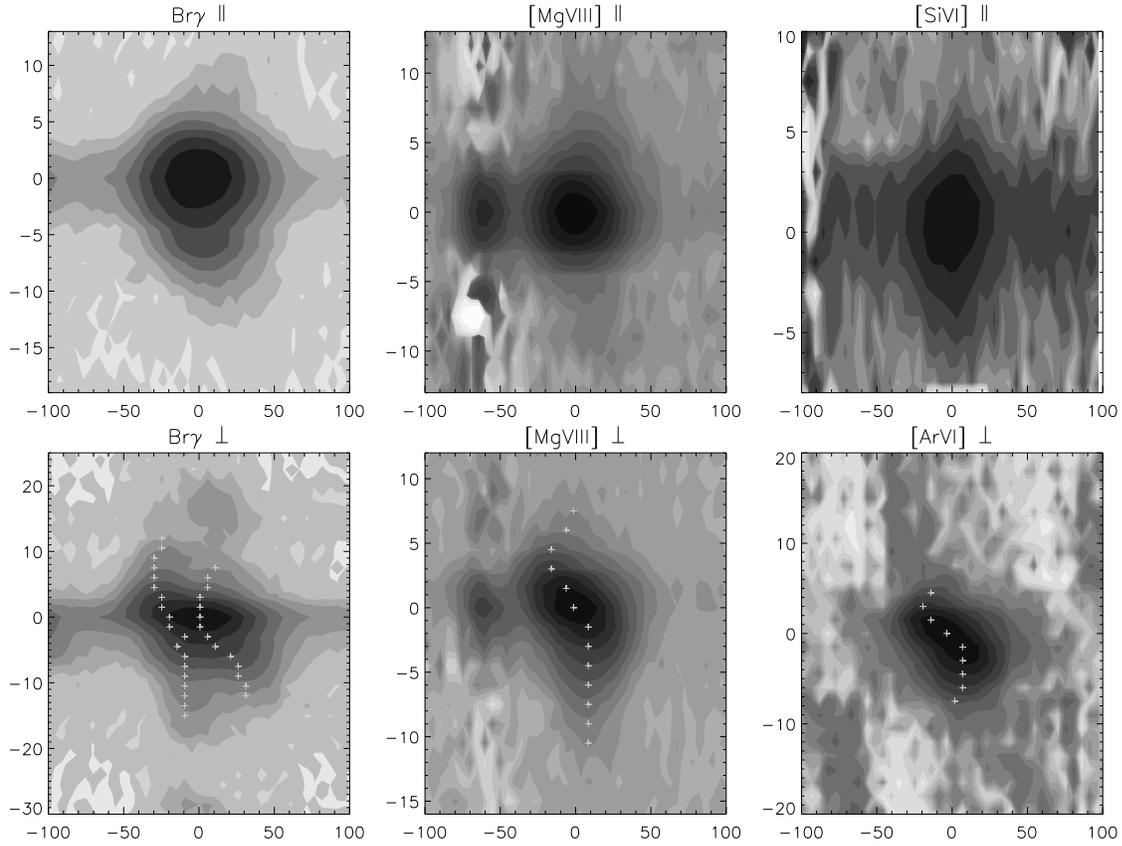}}
\end{center}
\caption{ Results of our echelle observations of NGC~6302.  The two
slit positions are referred to as $\parallel$ for the direction along
the waist of the nebula, and $\perp$ along the bipolar axis. From left
to right are shown Br$\gamma$, \hbox{[Mg\,{\sc viii}]} 3.028$\mu$m and
\hbox{[Ar\,{\sc vi}]} 4.528$\mu$m. The y-axis is the offset along the
slit in arcsec, and the x-axis shows the velocity in the object's rest
frame (km s$^{-1}$).  The contour levels are at the maximum flux
density times powers of 1/2. The crosses correspond to the maxima for
each row. East is towards positive offsets in the frames with the
$\perp$ slit orientation.}\label{fig:echelle}
\end{figure*}

The velocity profiles along the echelle slit are shown in Figure
\ref{fig:echelle} for NGC~6302. A radically different velocity
structure can be seen when comparing the coronal ions to Br$\gamma$
for the $\perp$ slit orientation, along the bipolar axis. The ions
show an outflow type of profile, whereas Br$\gamma$ displays the
characteristic X-shaped profile obtained for optical recombination
lines, and for [N\,{\sc ii}] by Elliot \& Meaburn (1977).  This
difference can be checked at offset --8$''$, where the double gaussian
fit to the Br$\gamma$ line profile has clearly separated centroids,
whereas \hbox{[Mg\,{\sc viii}]} or \hbox{[Ar\,{\sc vi}]} are centrally
peaked (the maxima at each offset are shown in crosses). The
$\parallel$ slit orientation shows narrow and centrally peaked
velocity profiles, in particular for \hbox{[Mg\,{\sc
viii}]}. Br$\gamma$ is less extended in the $\parallel$ direction,
consistent with the structure expected from a bipolar flow in the
$\perp$ direction.

\begin{figure*}
\begin{center}
\resizebox{16cm}{!}{\epsfig{file=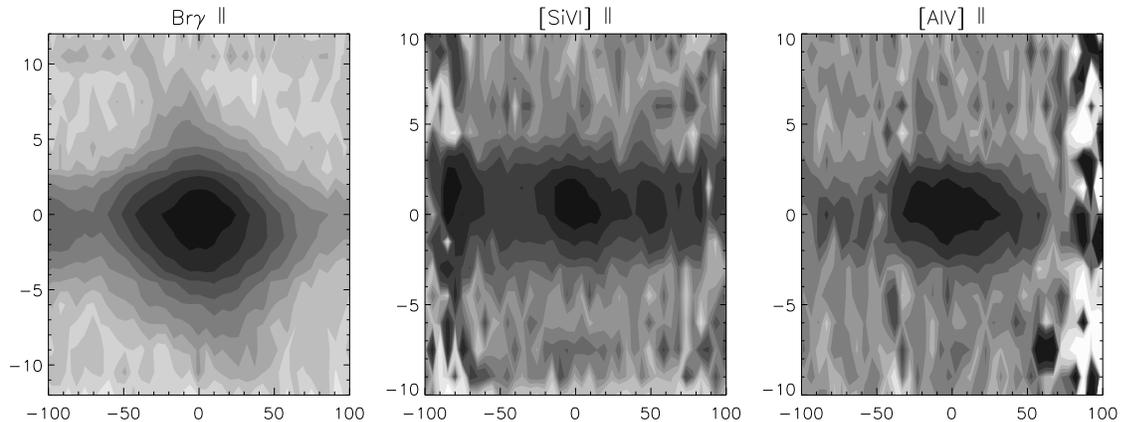}}
\end{center}
\caption{ Results of our echelle observations of NGC~6537. From left
to right are shown Br$\gamma$, \hbox{[Si\,{\sc vi}]} 1.963$\mu$m and
\hbox{[Al\,{\sc v}]} 2.883$\mu$m.  The slit was oriented along the
waist of the nebula.  The y-axis is the offset along the slit in
arcsec, and the x-axis shows the velocity in the object's rest frame
(km s$^{-1}$).  The contour levels are at powers of (1/2) times the
maximum flux density for each frame.}\label{fig:frames_n6537}
\end{figure*}

For NGC~6537, velocity profiles with the $\parallel$ slit orientation
only are available. Figure \ref{fig:frames_n6537} shows that they have
a similar structure to the $\parallel$ slit orientation in
NGC~6302. The lines are also quite narrow.

Table \ref{table:Vfwhm} lists the widths of the lines (FWHM, the
instrumental line widths were subtracted in quadrature) obtained by
fitting the spectra extracted from the collapsed slits. The
uncertainties stem from the instrumental line-widths. The line-widths
from the arc lamp spectra for the two [Mg~{\sc viii}] and the [Al~{\sc
v}] echelle frames are consistent with $\Delta$V=26$\pm$1~km\,$s^{-1}$
over the central third of the array, although the [Ar~{\sc vi}] arc
frame had a 20~km\,s$^{-1}$ wide line in the centre. But it is clear
that the lines are very narrow indeed, and might even not be resolved
in the case of \hbox{[Si\,{\sc vi}]} in NGC~6537.

\begin{table}
\caption{ FWHM line-widths $\Delta$V in km\,s$^{-1}$, together
with the ionization potential (IP, eV) required to produce the
observed ion, and the wavelength $\lambda$ in $\mu$m.}\label{table:Vfwhm}
\begin{tabular}{l|rrrrr} \\ \hline
           	           &   Br$\gamma$ & \hbox{[Mg\,{\sc viii}]} & \hbox{[Si\,{\sc vi}]} & \hbox{[Ar\,{\sc vi}]}       & \hbox{[Al\,{\sc v}]}            \\
$\lambda$ [$\mu$m]         & 2.166 &  3.028   & 1.963  & 4.528        & 2.883            \\
IP [eV]                    & 13.6  &  225     & 167    & 75.0         & 154              \\  \hline
\multicolumn{2}{l}{NGC 6302}     &&&&                                                             \\
$\Delta$V\,$\parallel$                &  31(5)&  18(2)    & 22(5)   &  ---         & ---           \\
$\Delta$V\,$\perp$                    &  41(5)&  24(2)    & ---     & 27(5)         & ---           \\   \hline
\multicolumn{2}{l}{NGC 6537}     &&&&                      \\
$\Delta$V\,$\parallel$                &  39(5) & ---      & 14(10)  & ---          &30(20)            \\
\end{tabular}
\end{table}


\begin{figure}
\begin{center}
\resizebox{8cm}{!}{\epsfig{file=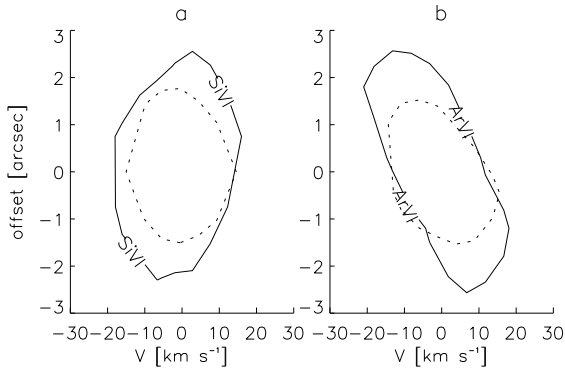}}
\end{center}
\caption{ Spatial and velocity extent of \hbox{[Mg\,{\sc viii}]}
(dotted line), \hbox{[Si\,{\sc vi}]} and \hbox{[Ar\,{\sc vi}]}. The
contours are taken at half maximum a) for the slit oriented along the
dusty lane, b) for the slit oriented along the bipolar axis. The
registration of the contours is uncertain to $0.1-0.2''$.}
\label{fig:strat}
\end{figure}

\begin{figure}
\begin{center}
\resizebox{8cm}{!}{\epsfig{file=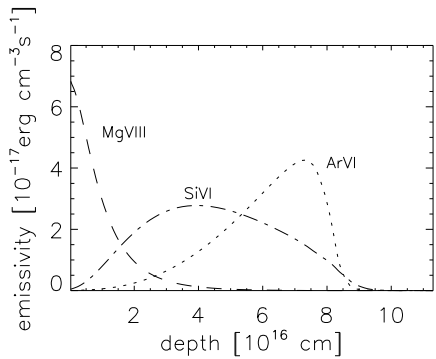}}
\end{center}
\caption{The ionization structure of a uniformly dense nebula as
computed by {\footnotesize CLOUDY}, with NGC~6302's parameters.}\label{fig:cloudy_struct}
\end{figure}

The velocity profiles give evidence for stratification in NGC~6302, as
can be seen in Figure \ref{fig:strat}. The contour at half maximum
corresponding to \hbox{[Mg\,{\sc viii}]} (dotted line), at
IP$=224.95$~eV, is less extended than those of \hbox{[Ar\,{\sc vi}]},
IP$=75.04$~eV, and \hbox{[Si\,{\sc vi}]}, IP=166.77eV. The nebular
emission in each frame did not necessarily peak on the same row, which
affects the registration of the contours between different
observations. The relative positioning of the contours is also
affected by the limited spatial resolution of about 1.5$''$ along the
slit, and by the velocity resolution of $\sim$10\,km\,s$^{-1}$ per
pixel along the dispersion axis. It can be noted that the lower
ionization species are more spatially extended than \hbox{[Mg\,{\sc
viii}]}, which matches the structure of a photoionized nebula, shown
in Figure \ref{fig:cloudy_struct}. The velocity width of the lines
also show stratification, implying that material is accelerated
outwards from the nucleus.


\subsection{The absence of a wind blown cavity}

Most kinematical studies of bipolar nebulae have been based on optical
echelle spectroscopy of [N\,{\sc ii}] 6584~\AA , which does not probe
the high excitation regions and suffers from extinction. The general
structure that emerges from the optical studies is typically a bipolar
cavity surrounded by an expanding H\,{\sc ii} region, which in turn is
confined by some higher density medium (sometimes detected in H\,{\sc
i} or CO). The observational results are usually compared with the
interacting stellar winds model (ISW, Kwok et al. 1978), with fairly
good agreement. For instance, Corradi \& Schwarz (1993) studied
NGC~6537 and Hb\,5 in [N\,{\sc ii}] and compared their results with the
ISW model of
Icke et al. (1989); NGC~6537 has also been studied in [N\,{\sc ii}] by
Cuesta et al. (1995) and its kinematical structure was consistent with
refraction of a fast wind on the walls confining an evacuated cavity
(model for Herbig-Haro objects from Cant\'{o} 1980); Bryce et
al. (1996) established the velocity structure of NGC 650-1 using
[N\,{\sc ii}] and [O\,{\sc iii}], and the results were in reasonable
agreement with the GISW models presented in Frank and Mellema
(1994). A common feature in the models quoted previously is that a hot
evacuated cavity ($N_\mathrm{e} \sim 10~$cm$^{-3}$) occupies a good portion of
the PN's volume, whose walls are expanding with velocities of order
50~km\,s$^{-1}$.

In the case of NGC~6302, the above picture would seem to be confirmed
by the CGS4+echelle observations of Br$\alpha$. Fig. \ref{fig:echelle}
shows that the Br$\alpha$ profiles double-peak further out than 8$''$
from the nucleus, which is an indication for a possible decrease in
proton density along the bipolar axis. This could also be extrapolated
from the optical echelle measurements of recombination lines (Elliott
\& Meaburn 1977), and from the H$\beta$ image of Ashley (1990), under
the assumption of axial symmetry.

But the use of coronal ions as tracers of the velocity field produces
quite a different picture.  The observed ionization stage
stratification is consistent with a {\small CLOUDY} model for a nebula
at constant density (Figure \ref{fig:cloudy_struct}). We also have no
indications for a torus type of structure rather than a disk around
the waist of the nebula, at least at about 1$''$ resolution, as can be
seen from the frames along the dusty lane. The same situation seems to
apply to NGC~6537 (Figure \ref{fig:frames_n6537}). Additionally, the
single peaked profiles of the coronal lines out to 8$''$ offsets along
the bipolar axis (in NGC~6302, Fig. \ref{fig:echelle}) show that the
corresponding physical conditions cannot be that of a rarefied
plasma. The detection of coronal line emission requires $N_\mathrm{p}
\gs 10^{3}$\,cm$^{-3}$ out to 10$''$ offsets, and even higher if the
decrease in ionic abundance is taken into account\footnote{Using
detailed balance it is possible to determine the range in temperature
and density where the predicted ratio of the flux at nebular
conditions ($N_\mathrm{e} = 18000$~cm$^{-3}$ and $T_\mathrm{e} =
20000$~K, Section \ref{sec:abundances}) to the flux at
$(N_\mathrm{e},T)$ would put it below the noise level. A drop in
[Mg\,{\sc viii}] flux by a factor 1000 corresponds to a drop in
density from nebular conditions (Section \ref{sec:abundances}) to
$N_\mathrm{p} \geq 8\times10^{2}$~cm$^{-3}$. The limiting case
$N_\mathrm{p} = 8\times10^{2}$~cm$^{-3}$ corresponds to
$T_\mathrm{e}=10^{4}$~K.}.


Arguments against the existence of a sizeable evacuated cavity may
also be extrapolated from the VLA 6\,cm map of NGC~6302's central
20$''$ by Gomez et al. (1993). The 6\,cm continuum traces the emission
measure across the central regions of the nebula (provided
$T_\mathrm{e}$ is fairly uniform).  Under the assumption of axial
symmetry, the possibility of an extreme decrease in proton density from
nebular
conditions to $\sim$10~cm$^{-3}$ along the bipolar axis can be discarded
(note that for a given 6\,cm flux density, an increase in
$T_\mathrm{e}$ requires an increased $EM$).  The CGS4+echelle results
reported here confirm the filled-in picture evident from the VLA map,
this time using tracers of highly excited gas. Also early observations
of PNe by Wilson (1950) showed that the central `cavities' in the long
slit
spectra of low-ionization species (such as [O\,{\sc ii}] and [N\,{\sc
ii}]) are filled-in, spatially and spectrally (at $\sim$10~km\,s$^{-1}$
resolution), by higher-ionization species such as [Ne\,{\sc v}].
Wilson (1950) thus reported a stratification in ionization potential,
including the intermediate ionization species [O\,{\sc iii}] and
[Ne\,{\sc iii}].

NGC~6302 has previously been studied in \hbox{[Ne\,{\sc v}]} by
Meaburn \& Walsh (1980). They concentrated on the detection of a fast
wind in the broad wings of \hbox{[Ne\,{\sc v}]}, which are absent in
their [N\,{\sc ii}] profiles.  This discovery has been interpreted by
Barral et al.  (1982) in terms of the Cant\'{o} (1980) model for H-H
objects.  In this model, flows are generated as the stellar wind
refracts on a dense disk surrounding the star and is accelerated along
the walls of the bipolar nebula. The shape of the walls is determined
by equating the ram pressure of the wind to that of the surrounding
medium.  The conclusion of Barral et al. (1982) within this model is
that the dominant excitation mechanism is still radiative
ionization. In this case, an altogether different model for
photoionization should be considered, in which the nebula has in fact
a shell structure, the hot bubble consisting only of the stellar wind
itself.  But the spectra of \hbox{[Ne\,{\sc v}]} published by Meaburn
\& Walsh do not show a double peaked structure, at a resolution of
about 10~km\,s$^{-1}$.  Although Meaburn \& Walsh (1980) interpreted
the geometry of NGC~6302 as an evacuated cavity, with \hbox{[Ne\,{\sc
v}]} emission originating from the walls, their results do not show
the corresponding velocity profile. We believe a filled-in structure
is not inconsistent with observing different velocity components,
provided the nebula has a clumpy structure.

This Section points towards the absence of a wind-blown cavity filled
with a hot rarefied plasma, in the case of the highest excitation PN
known.  We suggest that the clumpiness of the nebula precludes the
formation of a wind-blown cavity. This point should be further
investigated to see under which conditions, if any, bipolar PNe would
show signs of a hot bubble.  Similar conclusions have been reached by
O'Dell (1998) and Henry et al. (1999) in the case of the Helix Nebula 
(NGC
7293), namely that the ionized phase of the nebula is a disk rather than a
ring, filled by highly ionized gas. The current ISW
dynamical models that account for the shaping of PNe have also been
used to explain a variety of astrophysical objects, and even the
collimation of jets in AGN (Icke et al. 1992) and young stellar
objects (Mellema \& Frank 1997). But the ISW model was built from and
checked against optical studies of relatively low ionization species
in PNe, and we have shown here that the observational information should
be extended by studies of the cores of other nebulae in
higher-excitation species.

\subsection{What are the physical conditions in the fast wind?}

The signal to noise ratio of our echelle observations of NGC~6302 in
\hbox{[Ar\,{\sc vi}]} and \hbox{[Mg\,{\sc viii}]} is higher than that
of \hbox{[Ne\,{\sc v}]}\,3426~\AA~ presented by Meaburn \&
Walsh. Also, \hbox{Ne\,{\sc v}} has an ionization energy intermediate
between that of Ar\,{\sc vi} and Mg\,{\sc viii}, while the IR coronal
lines and the optical [Ne~{\sc v}] line have similar critical
densities (at $10^4$~K, the IRON project collision strengths give
$N_\mathrm{crit}$([Ar\,{\sc vi}])$\sim~10^{6}$~cm$^{-3}$,
$N_\mathrm{crit}$([Ne\,{\sc v}])$\sim~5\times10^{6}$~cm$^{-3}$,
$N_\mathrm{crit}$([Mg\,{\sc viii}])$\sim~10^{7}$~cm$^{-3}$).  So why are
no extended wings observed in the IR coronal lines?


Without mass loading, the fast stellar wind should vary in density
with angular distance in arcsec, $\alpha$, from the central star
approximately as 10/$\alpha^2$ cm$^{-3}/''^{2}$ (for
$dM/dt=10^{-8}$~M$_{\odot}$yr$^{-1}$ and
$V_\mathrm{fw}=1000\,$~km\,s$^{-1}$). This would not be detected in IR
emission because of insufficient collisional excitation, leading to a
very low excited ion column density. But the physical conditions where
the broad wings originate could be intermediate between the fast
stellar wind and nebular conditions. Let us consider the implications
of assuming collisional excitation for the fast wind observed by
Meaburn \& Walsh (1980). For the \mbox{\small
$^{3}$P$_{2}$-$^{1}$D$_{2}$} \hbox{[Ne\,{\sc v}]}\,3426\AA~
transition, the line intensity can be written as

\begin{equation}
\label{eq:line}
I_{i-j}=\frac{1}{4\pi}{\int}\epsilon\, ds  \,b \,A_{ij} \, h {\nu}
\,n_\mathrm{up} \, N_{o},
\end{equation}

where b is the branching ratio, $\epsilon$ the filling factor,
$n_\mathrm{up}$ is the fraction of ions in the upper level, and
$N_{o}$ is the number density of ions. We have given a number of
justifications for the fast wind coexisting with the denser clumps,
which occupy a significant fraction of the volume ($\epsilon \sim 0.5$),
so the fast wind's filling factor should be similar. The ratio of the
intensity from the narrow component of the line to that from the fast
wind will be given approximately by
\begin{equation}
\label{eq:Rline}
\frac{I_\mathrm{narrow}}{I_\mathrm{wings}} \approx \frac{(n_\mathrm{up} N_\mathrm{p})_\mathrm{narrow}}
{(n_\mathrm{up} N_\mathrm{p})_\mathrm{wings}}
\end{equation}
as long as the ionic abundance does not change significantly between the
two components. Meaburn \& Walsh (1980) give
\begin{equation}
I_{\hbox{[Ne\,{\sc v}]}}^\mathrm{wings} \approx (1/3)
I_{\hbox{[Ne\,{\sc v}]}}^\mathrm{narrow} .
\end{equation}
If the ionic abundances were substantially different in the fast wind,
such an important contribution to the total line flux would put into
question photoionization models. A two-phase nebular gas should be
considered. Our IR observations do not show the fast wind, so the
photoionization models presented here should not be affected since we
use no optical lines. A detailed photoionization model including a
range of optical and infrared lines could constrain possible
variations in ionic abundances among the two phases. Nonetheless,
under the assumption that ionic abundances do not change between clump
and fast wind, the range of physical conditions in the fast wind that
gives a [Ne\,{\sc v}] line ratio of 1/3 is shown in Figure
\ref{fig:fwcond}, together with the corresponding ratio for the
case of \hbox{[Ar\,{\sc vi}]} and \hbox{[Mg\,{\sc viii}]}. The noise
levels reported in Figure \ref{fig:echelle_n6302} yield the following
upper limits on the broad wings fluxes in [Mg\,{\sc viii}] and
[Ar\,{\sc vi}], assuming the broad wings flux is uniformly
distributed over 800~km\,s$^{-1}$,
\begin{eqnarray}
I_{\hbox{[Mg\,{\sc viii}]}}^\mathrm{wings} &\ls & ~0.02~I_{\hbox{[Mg\,{\sc
viii}]}}^\mathrm{narrow} \\
I_{\hbox{[Ar\,{\sc vi}]}}^\mathrm{wings} & \ls & ~0.2~ I_{\hbox{[Ar\,{\sc vi}]}}^\mathrm{narrow}.
\end{eqnarray}
Using Figure \ref{fig:fwcond}, it can be concluded that broad wings should
also be apparent in \hbox{[Ar\,{\sc vi}]}, and especially in
{\hbox{[Mg\,{\sc viii}]}, for the acceptable values of densities and
temperatures. It seems that collisional excitation cannot account for
the broad wings observed in the optical, at least in these simple
terms. A more thorough study including a broader range of lines is
required.

\begin{figure}
\begin{center}
\resizebox{8cm}{!}{\epsfig{file=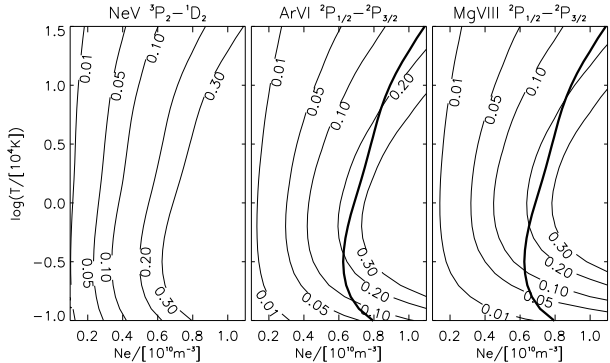}}
\end{center}
\caption{ Estimation of the ratio of the broad component to the narrow
component as a function of $\log(T/10^{4}$K) (ordinate) and
electron number density. The transitions correspond to \hbox{[Ne\,{\sc
v}]} 3426\AA~ and \hbox{[Ar\,{\sc vi}]} 4.528$\mu$m, and
\hbox{[Mg\,{\sc viii}]}~3.0284$\mu$m. The thick line limits the locus
of $\{\log(T/10^{4}$K$),N_\mathrm{e}\}$ that can account for the broad
wings observed in
\hbox{[Ne\,{\sc v}]} 3426\AA~.} \label{fig:fwcond}
\end{figure}

\section{Conclusions}
\label{sec:concl}

The high excitation PNe NGC~6302 and NGC~6537 have been studied with
medium and high resolution infrared spectroscopy.  We provide a new IR
estimate for the heliocentric radial velocity of NGC~6302, finding a
$4.1\pm2.7$~km\,s$^{-1}$ difference from optical estimates, possibly
due to extinction within the ionized region. The coronal line
wavelengths we derive are in agreement with those of Reconditi \&
Oliva (1993), but not with those of Feuchtgruber et al. (1997).

Our medium resolution spectra show \hbox{[Si\,{\sc ix}]}\,3.934$\mu$m
in emission, which is used with other IR coronal ions to confirm the
nebular excitation mechanism to be photoionization, corresponding to a
blackbody central star temperature of about 250\,000~K for
NGC~6302. The only anomalous elemental abundance derived from our
NGC~6302 spectra is that of aluminium, found to be depleted relative
to solar by a factor of more than one hundred, presumably due to being
locked in grains. The strong Al depletion implies that some dust is
well mixed with the highly ionized gas where the radiation field is
hard enough to produce Al\,{\sc vi}, at IP=154~eV.  A similar Al
depletion is inferred for NGC~6537. For the case of single star
evolution, the revised central star temperature for NGC~6302 is
consistent with post-AGB model tracks (Bl\"{o}cker 1995) for ZAMS
masses between 4 and 5 M$_{\odot}$.  NGC~6302's N/O and C/O gas phase
abundance ratios are also consistent with a progenitor main sequence
mass $\ls$ 5 M$_{\odot}$, from an adaptation of the synthetic AGB
models of Gronewegen \& deJong (1993).

Our IR echelle observations reveal that the infrared coronal lines are
very narrow and strongly peaked at the centre, with no evidence of
broad wings from fast winds.  The discrepancy with the
$\sim$1000~km\,s$^{-1}$ broad wings previously observed for 
\hbox{[Ne~{\sc v}]\,3426~\AA~}
cannot be explained by collisional excitation in a photoionized gas,
at least in a simplified treatment. Further observations are required
to confirm the \hbox{[Ne~{\sc v}]} broad wings, and to build detailed
models of the interaction between the fast stellar wind and the
nebula.  The narrowness of the line profiles also makes NGC~6302 and
NGC~6537 ideal objects for the study of fine-structure in coronal
ions. The high resolution observations show that the cores are
stratified in ionization stages, consistent with a photoionized nebula
of uniform density. At 8$''$ offset from the centre of NGC~6302,
Br${\gamma}$ shows double-peaked profiles, in agreement with previous
optical studies based on recombination lines and [N\,{\sc ii}]. In
contrast, the line profiles from the coronal ions are single-peaked at
the same offset. Thus our observations show no signs of a wind-blown
bipolar cavity. The nebula, although clumpy enough to accommodate
coexisting winds at very different velocities, seems to have a fairly
uniform density as close as 1$''$ to the nucleus. We suggest
that the clumpiness of the nebula precludes the formation of a
wind-blown cavity.


\section*{Acknowledgments}

We are very grateful to Tom Geballe for obtaining the CGS3 spectra, to
Gary Ferland for help with {\small CLOUDY}, and to Luc Binette, John
Dyson, Tom Hartquist and Melvin Hoare for helpful discussions. We are
also grateful to PATT for the time allocation on UKIRT, which is
operated by the Joint Astronomy Centre on behalf of the Particle
Physics and Astronomy Research Council. S.C. acknowledges support from
Fundaci\'{o}n Andes and PPARC through a Gemini studentship.

\appendix

\section{Fluxes and  accurate wavelengths for fine structure lines,
and the heliocentric velocities of NGC~6302 and
NGC~6537}\label{sec:acc}

A number of fine structure lines can be found in the spectrum of
NGC~6302 and NGC~6537, whose accurate wavelengths represent valuable
information and can be compared with previous measurements. The low
resolution spectra were calibrated using nebular H\,{\sc i} lines when
present near the lines of interest, otherwise using a quadratic fit to
Ar or Kr arc lamp spectra (deviations from a linear law were found to
be negligible). The calibration of the echelle observations is
described below. We found the heliocentric velocity of NGC~6302 to be
--34.8$\pm$1~km\,s$^{-1}$, using \hbox{H\,{\sc i}}~5-10 at
3.0384$\mu$m in the two \hbox{[Mg\,{\sc viii}]} frames. This
corresponds to --27.3$\pm$1~km\,s$^{-1}$ in the local standard of rest
(kinematic LSR), which agrees closely with the H76$\alpha$ measurement
of -29$\pm$2~km\,s$^{-1}$ by Rodriguez et al. (1985, their Fig. 2
shows that the observed peak is red shifted with respect to their
Gaussian fit by about 5~km\,s$^{-1}$). But the heliocentric velocity
of NGC~6302 obtained from optical studies (compilation by Durand et
al. 1998, based on Acker 1975) has been reported to be
--38.9$\pm$2.5~km\,s$^{-1}$.  On the other hand we measure a
heliocentric velocity for NGC~6537 of
\hbox{--17.8$\pm$3~km\,s$^{-1}$}, using \hbox{H\,{\sc i}}~5-11 at
2.872213$\mu$m on the [AlV] frame, which is in agreement with the
value listed in Durand et al. (1998) of
--16.9$\pm$1.9~km\,s$^{-1}$. It should be stressed that the long
wavelength side of \hbox{H\,{\sc i}}~5-11 was very noisy, the
heliocentric velocity we report for NGC~6537 assumes that the width of
the line is the same as for Br$\gamma$, but any error would result in
a higher (i.e. red shifted) velocity.

By adopting the estimated systemic velocities from the echelle
spectra, we can achieve consistent wavelength determinations for the
extinction insensitive infrared coronal lines. Considering the quoted
uncertainties in the case of NGC~6302, there could be a slight
discrepancy with the optical measurements, which is most likely due to
internal extinction within the ionized regions of the nebula. Huggins
\& Healy (1989) report a CO(2-1) centroid for NGC~6302 of
$V_\mathrm{lsr}$~=~--40~km\,s$^{-1}$, blue shifted in excess of
10~km\,s$^{-1}$ from the measurements reported here.  The CO(2-1)
profile is double peaked, and the approaching side has about twice the
peak brightness temperature as the receding side (we conducted SEST
observations of $^{13}$CO(2-1) which show the same profile, and that
shall be presented in a forthcoming article), hinting at possible
radiative transfer effects. However, we do not base any of the
following on the mm-line centroids, because the line of sight towards
NGC~6302 crosses cold galactic molecular emission. 


Table \ref{table:lambdas} gives a summary of our measurements, where
wavelengths are given in air and in the nebular rest frames (the
refractive index of standard air as a function of wavelength was taken
from Edl\'{e}n 1953). The uncertainty in the calculated wavelengths is
dominated by the accuracy of the adopted heliocentric velocities, in
the best cases.  Below we provide some notes and compare our
measurements with those found in the literature, where available.

\begin{table*}
\caption{Rest wavelengths (in air) for emission lines  from NGC~6302,
or from NGC~6537 when specified. The numbers in brackets are the
uncertainties in the last decimal places.  Fluxes are given for the
whole slit, without correction for extinction or blends and with a
typical uncertainty of 20 per cent. The slit orientation for each line
can be inferred from the tabulated fluxes.}\label{table:lambdas}
\begin{tabular}{ccccccccc}\hline
       Ion  &       $ i<j $ &   $\lambda$ [$\mu$m] &   $\lambda$ [$\mu$m] &  \multicolumn{2}{c}{$\lambda$ [$\mu$m]~$^{b}$}& \multicolumn{3}{c}{Flux [W m$^{-2}$]}    \\
            &               &  	R=1500		   &  	R=20000		&   \multicolumn{2}{c}{(reference)}&  $\parallel$ 3'' slit   & $\parallel$ 1''slit&$\perp$1''slit\\ \hline
            &               &  			   &        		&	                           &	 	             &                      &            \\
Si\,{\sc vi}&$ ^{2} $P$_{3/2}$-$^{2}$P$_{1/2}$&           &  1.96287(10)           & 1.96287(10)	& (3)& 			    &$1.6{\times}10^{-16}$&                       \\
Si\,{\sc vi}~$^{c}$ &$ ^{2}$P$_{3/2}$-$^{2}$P$_{1/2}$&         &  1.96311(20)           & 1.96287(10)	& (3)& 	                    &$1.0{\times}10^{-17}$&                       \\
Al\,{\sc v}~$^{c}$  &$ ^{2}$P$_{3/2}$-$^{2}$P$_{1/2}$&  	 &  2.882975(19)	& 2.90440(37) 	& (1)&                      &$9.6{\times}10^{-18}$&                       \\
Mg\,{\sc viii}      &$ ^{2}$P$_{1/2}$-$^{2}$P$_{3/2}$& 3.02788(67)& 	3.027661(20)	& 3.02713(25)	& (3)&$	2.9{\times}10^{-15}$&$1.7{\times}10^{-15}$&                       \\
(second)    ~$^{a}$&                              &          &      3.027033(20)    &               &    &                      &$1.9{\times}10^{-16}$&                       \\
Mg\,{\sc viii}      &$ ^{2}$P$_{1/2}$-$^{2}$P$_{3/2}$&           &	3.027648(20)	& 3.02713(25)	& (3)&                      &                     & $1.4{\times}10^{-15}$ \\
(second)   ~$^{a}$&            &                            &      3.027003(20)    &               &    &                      &                     & $7.3{\times}10^{-17}$ \\
K\,{\sc vii}       & $ ^{2}$P$_{1/2}$-$^{2}$P$_{3/2}$&  3.1899(21)& 			& 3.18966(15) 	& (1)&$	4.2{\times}10^{-16}$&                     &                       \\
Al\,{\sc vi}       &$ ^{3}$P$_{2}$-$^{3}$P$_{1}$    &  3.65952(81)&			& 3.661(14)	& (2)&$	2.1{\times}10^{-16}$&                                             \\
Si\,{\sc ix}       &$ ^{3}$P$_{0}$-$^{3}$P$_{1}$    & 3.9346(26)& 			& 3.935 	& (4)&$	3.5{\times}10^{-17}$&                     &                       \\
Ar\,{\sc vi}       &$ ^{2}$P$_{1/2}$-$^{2}$P$_{3/2}$&  4.5265(30) &  4.52799(15)& 4.52829(31)	& (1)        &$	8.7{\times}10^{-14}$&                     & $2.5{\times}10^{-14}$ \\
Na\,{\sc vii}      &$ ^{2}$P$_{1/2}$-$^{2}$P$_{3/2}$& 4.6908(31) &	 		& 4.68344(33) 	& (1)&$	7.4{\times}10^{-15}$&                     &                       \\ \hline
\end{tabular}
\medskip

$^{a}$ Unidentified secondary line in the \hbox{[Mg\,{\sc viii}]} 3.028$\mu$m frame.\\
$^{b}$ Previous measurements from  1), Feuchtgruber et al. (1997, corrected to values  in air); 2, Greenhouse et al. (1993); 3), Reconditi \& Oliva (1993); 4), Oliva et al. (1994).\\
$^{c}$ Measured from NGC~6537.\\
\end{table*}

[Mg\,{\sc viii}]: We have two echelle measurements corresponding to
the two slit orientations. A comparison of the standard star spectra
with a high-resolution atmospheric transmission model (based on the
HITRAN database, Rothman et al. 1992) gave an initial quadratic fit to
the dispersion law, which we used to identify Argon lines from the arc
lamp frames. The arc lamp calibration agrees very well with the
atmosphere calibration, although with shifts of order 10~km\,s$^{-1}$
towards the edges of the array, which probably stem from the standard
star not being precisely centred within the 1'' slit. We used the arc
lamp to calibrate the \hbox{[Mg\,{\sc viii}]} frame.  Many lines were
identified as leaks from neighbouring orders, i.e. the Argon arc lamp
spectra in order 36 also showed lines from orders 37 and 38, implying
that the CVF blocked wavelengths outside of about 0.1$\mu$m from the
CVF wavelength of 1.515\,$\mu$m. The calibrated dispersion axis is
strongly non-linear, with deviations from a linear law of up to
20~km\,s$^{-1}$, and offset from the estimated scale by
--300~km\,s$^{-1}$ to --50~km\,s$^{-1}$ from side to side of the
array. We estimate the uncertainty in the \hbox{[Mg\,{\sc viii}]}
wavelength to be 2~km\,s$^{-1}$, dominated by the relative velocity of
NGC~6302. H\,{\sc i} 5-10 is on the same frame, and allows an accurate
estimate of the doppler shift from NGC~6302. The low resolution spectrum
estimate is accurate to better than one sample element
(i.e. R=1500$\times$3), but the secondary line is blended.

[Ar\,{\sc vi}]: The echelle wavelength calibration was based on a
xenon lamp, which showed one arc line only, at 2.2618283$\mu$m (Outred
1978) in second order, very close to the centre of the array. Because
of the modulation due to the CVF fringes it was impossible to use the
standard star spectrum to calibrate the wavelength dispersion. We
used the quadratic dispersion law for \hbox{[Mg\,{\sc viii}]}
3.03~$\mu$m, shifted to match the xenon arc
line. The uncertainty should be about 10~km\,s$^{-1}$. The low
resolution spectrum did not provide a satisfactory measurement because
the line was strongly affected by random bad pixels, with an accuracy
of about one resolution element.

[Al\,{\sc v}]: Here the echelle spectrum was of NGC~6537, rather than
NGC~6302. We calibrated the dispersion axis by the same procedure as
for \hbox{[Mg\,{\sc viii}]}, and confirm that the CVF does not block
wavelengths within 0.1~$\mu$m from the CVF wavelength. 
\mbox{H\,{\sc i} 5-11} 2.872213$\mu$m was also
present in the same frame, although it is noisy this line allows an
estimate of the systemic velocity of NGC~6537. No measurements are
provided for the low resolution spectrum because the line is blended
with H\,{\sc i} 5-11, which represents about 75 per
cent of the flux.

[Si\,{\sc vi}]: we used the airglow OH line at \mbox{1.95878 $\mu$m}
(\mbox{1.95932 $\mu$m} in vacuum from the UKIRT worldwide
webpages). The line is very faint and was discernible across the
array only after co-adding all the object frames, and there is an
uncertainty of about 10~km\,s$^{-1}$ in the position of the line. The
standard star spectra seem to be affected by local absorption related
to the instrument, which is also apparent in the nebular continuum
emission from both PNe. The solar spectrum is at odds with the
standard's, so unfortunately the atmosphere cannot be used for
calibration. We adopted the dispersion scale from the \hbox{[Mg\,{\sc
viii}]} frame shifted to match the OH line. An uncertainty of no less
than 10~km\,s$^{-1}$ is expected in the dispersion scale, and the
\hbox{[Si\,{\sc vi}]} wavelength is thus accurate to about
15~km\,s$^{-1}$.

[Al\,{\sc vi}]: Only a low resolution wavelength determination is
available, but it should be reliable because that spectrum showed many
other lines, which all match the reference wavelengths. However the
best accuracy achievable is probably not better than one sample
element, or 67~km\,s$^{-1}$ (R~=~1500 oversampled 3 times),
because the line is noisy and lies near a zone of strong atmospheric
absorption.

[Si\,{\sc ix}]: this line was calibrated using nearby H\,{\sc i} Hu15,
but it is faint and the accuracy is only one resolution element. The
resulting wavelength is in good agreement with the measurement from
the coronal spectrum of the Circinus galaxy by Oliva et
al. (1994). This comes as a confirmation for their adopted radial
velocity since in Circinus the narrow line region is blue shifted
relative to the systemic velocity, and \hbox{[Si\,{\sc ix}]} is about
300~km\,s$^{-1}$ broad.

Finally, \hbox{[K\,{\sc vii}]} was calibrated with respect to an Argon arc
lamp and \hbox{[Na\,{\sc vii}]} with respect to the nearby \hbox{He\,{\sc
ii}} 6-7 line, although both were quite noisy and so the wavelengths are
estimated to be accurate to only one resolution element. 

The low resolution and the echelle measurements are consistent, and
agree in general with other published results, except for the {\it
ISO} measurements of Feuchtgruber et al. (1997). The agreement with
Reconditi and Oliva (1993) is improved when using the optical
heliocentric velocity of NGC~6302 discussed above.

The fluxes for emission lines in the CGS3 spectra are listed in Table
\ref{table:cgs3}, together with the {\it ISO} fluxes from Pottasch \&
Beintema (1997) and Pottasch et al. (1996).  In order to compare the
CGS3 and {\it ISO} fluxes, correction factors need to be applied to
account for the different beam sizes, which can be estimated using the
spatial information from the low resolution CGS4 spectra. In the case
of NGC~6302, at a 3$''$ pixel size the average spatial full width at
half maximum (FWHM) of HI lines along the slit was 2.8 pixels, and all
coronal ions had approximately the same width of 2 pixels.  For
NGC~6537, the HI FWHM was 1.7 pixels, and the coronal ions covered
approximately 1.3 pixels along the slit. These results suggest that,
for the purpose of scaling fluxes to the central 3$''$, all ions can
be approximated to have the same spatial extent. At the higher spatial
resolution of the echelle observations (see Figure \ref{fig:strat})
the FWHM for \hbox{[Mg\,{\sc viii}]}, \hbox{[Si\,{\sc vi}]} and
\hbox{[Ar\,{\sc vi}]} are about 3$''$, 4$''$ and 5$''$, and ions with
ionization potentials $\sim$100\,eV would have a spatial extent
intermediate between Si\,{\sc vi} and Ar\,{\sc vi}. The fluxes from
large beam observations should thus be about 2--3 times higher than
from the central 3$''$. This seems reasonable for $[$Mg\,{\sc
viii}$]$~3.028\,$\mu$m, listed in Table \ref{table:lambdas}, for which
Pottasch et al. (1996) give a flux of
3--6$\times$10$^{-15}$~W\,m$^{-2}$. For NGC~6302, fluxes through the
circular 4$''$ aperture of CGS3 should thus be 2--3 times less than
the fluxes for the whole nebula, assuming the emission lines have
uniform surface brightness over the central square 5--6$''$. From
Table \ref{table:cgs3}, a correction factor of $\sim$ 2.5 matches
\hbox{[Na\,{\sc vi}]~8.62\,$\mu$m} and \hbox{[S\,{\sc
iv}]~10.52\,$\mu$m}. We attribute the larger correction factor
required for \hbox{[Ne\,{\sc ii}]~12.8\,$\mu$m} to a larger spatial
extent, but the \hbox{[Ne\,{\sc vi}]~7.64\,$\mu$m} fluxes are
discrepant. \hbox{[Ne\,{\sc vi}]} lies on the edge of the atmospheric
window, and is likely to be affected by unresolved atmospheric
transmission features.

\begin{table}
\caption{Fluxes of emission lines in the 8--13$\mu$m spectra of NGC~6302
and NGC~6537 (in 10$^{-14}$~W\,m$^{-2}$), and comparison with {\it ISO}
measurements.}\label{table:cgs3}
\begin{tabular}{llccc}  \hline
                &  IP & \multicolumn{2}{c}{NGC~6302}  &  NGC~6537 \\
                &  $[$\,eV\,$]$ & CGS3      &    {\it ISO}           &   CGS3     \\
$[$Ne\,{\sc vi}$]$~7.64\,$\mu$m & 126.1 &  81.9  & 63.0      &  13.5       \\
$[$Ar\,{\sc v}$]$~7.90\,$\mu$m &  59.8  & 1.18    & --          &  0.86          \\
$[$Na\,{\sc vi}$]$~8.62\,$\mu$m &  138.4 & 0.56   & 1.1       &  0.23          \\
$[$Ar\,{\sc iii}$]$~8.99\,$\mu$m &  27.6  & 1.76  & --          &  0.80          \\
$[$S\,{\sc iv}$]$~10.5\,$\mu$m  & 34.8  & 4.77   & 8.5$^{a}$ &  6.77        \\
$[$Ne\,{\sc ii}$]$~12.8\,$\mu$m & 21.6  & 3.23   & 14.7$^{a}$&  1.14         \\
$[$Ar\,{\sc v}$]$~13.1\,$\mu$m &  59.8 & 0.56    & 2.9       &  0.61          \\
\end{tabular}

\medskip

$^{a}$ from Pottasch \& Beintema (1997), the other {\it ISO} fluxes are from
Pottasch et al. (1996).
\end{table}

\bsp
\label{lastpage}
\end{document}